%% file: main.tex
\documentclass[preprint,12pt,times]{elsarticle}

\usepackage[a4paper,margin=22mm]{geometry}
\usepackage{amsmath,amssymb,bm,mathtools}
\usepackage{graphicx}
\usepackage{booktabs,tabularx,array,multirow}
\usepackage{algorithm}
\usepackage{algpseudocode}
\usepackage{caption}
\usepackage{microtype}
\usepackage{placeins}
\input{landscapefig}
\usepackage{seqsplit}
\usepackage{xurl}
\usepackage{xcolor}
\usepackage[colorlinks=true,linkcolor=black,citecolor=black,urlcolor=blue,pdfborder={0 0 0}]{hyperref}
\usepackage[capitalise,nameinlink]{cleveref}
\makeatletter\providecommand{\theHALG@line}{}\renewcommand{\theHALG@line}{\thealgorithm.\arabic{ALG@line}}\makeatother
\crefname{figure}{Fig.}{Figs.}
\Crefname{figure}{Figure}{Figures}
\crefname{table}{Table}{Tables}
\crefname{equation}{Eq.}{Eqs.}
\Crefname{equation}{Equation}{Equations}
\crefname{section}{Section}{Sections}
\crefname{algorithm}{Algorithm}{Algorithms}
\crefname{appendix}{Appendix}{Appendices}
\DeclareUrlCommand\path{\urlstyle{same}}
\DeclareCaptionLabelFormat{continued}{#1~#2 (continued)}



\newcolumntype{Y}{>{\raggedright\arraybackslash}X}
\newcolumntype{Z}{>{\centering\arraybackslash}X}
\newcommand{\Rey}{\mathrm{Re}}
\newcommand{\dg}{\ensuremath{^{\circ}}}
\algrenewcommand\algorithmicrequire{\textbf{Input:}}
\algrenewcommand\algorithmicensure{\textbf{Output:}}

\journal{Journal of Computational Physics}
\begin{document}
\hypersetup{colorlinks=true,linkcolor=black,citecolor=black,urlcolor=blue,pdfborder={0 0 0}}
\begin{frontmatter}
\title{Task-preserving neural segmentation of overlapping shocks and vortex cores in compressible flows}
\author[aff1]{Ehsan Roohi\corref{cor1}}
\ead{roohie@umass.edu}
\cortext[cor1]{Corresponding author.}
\address[aff1]{Mechanical and Industrial Engineering, University of Massachusetts Amherst, 160 Governors Dr., Amherst, Massachusetts 01003, USA}

\begin{abstract}
Shock fronts and vortex cores often coexist and overlap in compressible flows, where walls, wakes and shear layers also produce strong gradients. Refining a network to detect one structure can therefore degrade its prediction of the other without revealing the loss in the optimized score. We present a task-preserving formulation for simultaneous shock and vortex-core segmentation. The experiments use two solvers and include supersonic diamond-airfoil flows at several incidences and Reynolds numbers, together with circular- and elliptical-cylinder flows. A shared primitive encoder feeds separate shock and vortex decoders with independent sigmoid outputs, allowing the two classes to overlap. Adaptation is confined to the relevant branch: zero-initialized adapters supply rotational diagnostics only to the vortex decoder, while corrected shock supervision updates only the shock decoder. All dependencies of the protected output remain fixed, and bitwise equality is verified on every evaluation field.
Compression and conservation-jump (Rankine--Hugoniot) signatures provide weak shock supervision; rotation and topology provide vortex candidates. Analytical oblique-shock rays, Billig's bow-shock correlation and an isentropic vortex supply references independent of these labels. Under the same corrected-target budget, the restricted model and a capacity-matched shared-decoder U-Net obtain comparable shock agreement. Their airfoil vortex-core Dice overlap scores, however, are 0.83 and 0.08, respectively. A soft retention penalty recovers most of the U-Net's lost core agreement. The frozen shock-adapted models locate the tested analytical oblique-shock rays within 0.003 chord. An additional branch identifies expanding-flow regions, which are distinguished from centred Prandtl--Meyer expansion fans through comparison with ideal shock--expansion theory.
\end{abstract}

\end{frontmatter}
\section{Introduction}
\label{sec:intro}
Shock waves and vortices govern many features of high-speed external and internal flows. Locating them automatically supports adaptive computation, reduced-order analysis, flow-regime identification, feature tracking and data-driven closures \cite{ref41,ref43}. The main difficulty is distinguishing structures with similar local signatures. A compressive discontinuity must be separated from wall edges, entropy layers, shear layers, expansion regions and numerical ringing. At the same time, a vortex may cross a shock, whereas strong wall vorticity need not indicate a coherent core. The deformation of both structures during shock--vortex interaction \cite{ref45,ref46} makes this distinction particularly demanding for a joint detector.

Shock detection in computational fluid dynamics (CFD) post-processing is commonly based on local sensors. Gradient sensors identify discontinuities but also respond to walls and contact surfaces. The normal-Mach-number criterion of Lovely and Haimes \cite{ref20} and the characteristics-based test of Kanamori and Suzuki \cite{ref21} improve specificity when reliable velocity fields are available. Dilatation-based sensors, including that of Ducros et al.\ \cite{ref3}, distinguish compression from expansion but remain broad across numerically thick fronts \cite{ref24}. Conservation-based tests require clean upstream and downstream samples \cite{ref2}. Fujimoto, Kawasaki and Kitamura \cite{ref1} combined Canny-type image-space edge detection \cite{ref4} with a Rankine--Hugoniot consistency test on Cartesian grids. Their sensor applies Canny and Sobel operators to pressure, followed by a gas-dynamic jump test, and was demonstrated on three inviscid and viscous CFD problems. Its directional operator, non-maximum suppression and jump criterion are interpretable and require no training. We use this sensor as a fixed classical reference without additional post-processing. It returns shock-edge cells, but does not represent front continuity, vortex cores or overlapping shock and vortex identities. In the separated Mach-3 field considered here, its output is strongly fragmented. Supplementary Section S2 examines the sensitivity of this result to the sensor parameters.

Learned shock detection has developed both as image analysis and as a component of numerical solvers. Image-based convolutional detectors have been trained on rendered flow fields \cite{ref22} and experimental shadowgraphs \cite{ref5,ref6}. Doroshchenko \cite{ref5} trained a You Only Look Once detector (YOLOv8 \cite{ref36}) on 1493 labelled shadowgraph images to identify vertical shocks, bow shocks, plumes and particles as bounding boxes. The study also tracked an oblique-shock angle using Canny edges and a Hough transform, and reported detection metrics on a held-out image split. These images came from one shock-tube facility; shocks were represented by boxes or fitted lines, with no vortex output. Within solvers, Ray and Hesthaven \cite{jcpRay2018} used an offline-trained multilayer perceptron to flag troubled cells in Runge--Kutta discontinuous Galerkin schemes. Beck et al.\ \cite{jcpBeck2020} trained convolutional shock detectors and localizers for discontinuous Galerkin methods, separating shock indication from stabilization. These approaches address different tasks from the dense, overlapping shock and vortex-core segmentation studied here.

Vortex identification is a different kind of problem. Vorticity measures local rotation but cannot tell a coherent core from a shear layer. The velocity-gradient-tensor criteria of Chong et al.\ \cite{ref26}, the $Q$ criterion, the $\lambda_2$ criterion of Jeong and Hussain \cite{ref7}, the swirling strength $\lambda_{ci}$ of Zhou et al.\ \cite{ref25} and the circulation-based $\Gamma_2$ function of Graftieaux et al.\ \cite{ref9} capture complementary aspects of rotation, strain dominance and angular coherence. Chakraborty et al.\ \cite{ref8} analyse how these criteria relate to one another, Haller \cite{ref27} discusses their frame dependence, and the visualization community has surveyed extraction methods at length \cite{ref30}. No single diagnostic survives every shock, wall, grid and sampling condition. A shock can displace the pressure minimum away from a real core, and a no-slip wall creates intense vorticity without a detached vortex. Learned vortex detectors include convolutional methods \cite{ref31,ref34} and VortexTransformer \cite{ref33}, which extracts vortices from material trajectories in unsteady two-dimensional flow. Because it works from material trajectories, VortexTransformer is not a comparator for the snapshot-based segmentation considered here, and its task differs from simultaneous shock and core segmentation in compressible flow; vortex-surface construction \cite{jcpXiong2017} addresses three-dimensional vorticity fields rather than two-dimensional fronts and cores. These limitations motivate a learned spatial decision informed by several diagnostics; a hard union or intersection of their masks cannot resolve every ambiguous structure.

Deep convolutional networks provide spatial context for connecting thin fronts and identifying compact cores. U-Net \cite{ref10} retains high-resolution information through skip connections, and independent sigmoid outputs allow it to represent overlapping classes. Our conventional U-Net baseline uses this formulation, so overlap representation is common to both architectures. A shared encoder also couples the task gradients, which can become negatively aligned \cite{ref37}. Loss re-weighting \cite{ref37,ref38} adjusts their relative contributions, while projecting conflicting gradients (PCGrad) \cite{ref11} removes conflicting components and retains compatible ones. Because transfer cannot be assumed under a change in input distribution \cite{ref56}, we evaluate changes in solver, Reynolds number, grid and incidence explicitly.

Freezing previously learned pathways is an established idea. Progressive neural networks \cite{rusu2016progressive} keep earlier columns while adding capacity for new tasks, and ControlNet \cite{zhang2023controlnet} trains conditioning pathways around a locked diffusion backbone. We apply the same principle to protected segmentation outputs. Exact preservation follows from unchanged computational dependencies, including inference-time state and operations, and we verify it with bitwise output tests. Whether the protected task is accurate is a separate question, and we assess it with its own references.

Joint shock and vortex-core segmentation requires more than an overlapping output representation. It also requires consistent class definitions, trajectory-level separation of training and evaluation data, and a way to measure interference between tasks. We use branch-restricted adaptation to control this interference: rotational diagnostics refine the vortex prediction without changing the shock branch, and revised shock supervision improves front coverage without changing the vortex branch. Exact preservation depends on keeping the full computational path of the protected output unchanged, which we check directly. Learned predictions, physics-only proposals and hybrid outputs are evaluated separately so that their respective contributions remain identifiable. Exact body geometry supplies an external domain constraint, and the frozen models are tested across solver, Reynolds-number, grid, incidence and transient-regime changes.

The formulation combines independent shock and vortex decoders with task-restricted physical augmentation. A freestream-referenced input representation enables controlled airfoil--cylinder comparisons while distinguishing local rotation, shear and compression. The numerical study evaluates spatial completeness, object recovery and numerical sensitivity alongside pixel agreement. It includes five learned configurations, three initialization seeds, cross-solver and Reynolds-number transfer, grid and time-step controls, and two geometric references independent of the training labels. U-Net, PCGrad and velocity-gradient diagnostics are established components; our contribution is their integration under explicit task-preservation constraints and the evaluation of what those constraints retain during adaptation.

Compared with the pressure-field sensor of Fujimoto et al.\ \cite{ref1} and the shadowgraph studies of Doroshchenko \cite{ref5} and Znamenskaya and Doroshchenko \cite{ref6}, the present formulation adds dense vortex-core predictions and retains overlap between the two structures. Each core has a probability field and an object identity; each shock has a centerline, an envelope and a boundary. These outputs support measurements of front localization, spatial continuity and task preservation that are outside the scope of edge-cell and bounding-box detection. The evaluation also extends beyond the three CFD demonstrations or single-facility image splits of those studies to controlled changes in Reynolds number, grid, solver, incidence and transient regime, using analytical and empirical shock references alongside label agreement.

The paper is organized as follows. \Cref{sec:problem} states the segmentation problem and the evaluation framework, \cref{sec:data} describes the flow cases and solvers, \cref{sec:method} develops the segmentation, adaptation and evaluation methods, \cref{sec:results} presents the transfer and controlled-comparison results together with the analytical and empirical reference checks, and \cref{sec:discussion,sec:conclusions} discuss the findings. Implementation details, extended results and provenance records are collected in the Supplementary material.

\section{Problem formulation and evaluation framework}
\label{sec:problem}
Let $\Omega_f$ denote the set of fluid cells of a two-dimensional field and $\Omega_s$ the exact solid domain occupied by the body. For each saved time $t$, the primitive-encoder formulation uses the canonical tensor
\begin{equation}
X(\mathbf{x},t)=\bigl[\log\rho,\ \log p,\ u,\ v\bigr](\mathbf{x},t),
\label{eq:input}
\end{equation}
where $\rho$ is the density, $p$ the pressure, and $(u,v)$ the Cartesian velocity components at position $\mathbf{x}=(x,y)$. The target ontology contains three classes: shock $S$, vortex core $V$, and background $B$. The classes $S$ and $V$ are not mutually exclusive, because a vortex can cross a shock; $B$ is the complement of $S\cup V$ after the regions marked as ignore have been removed. A geometry field $G$ and per-class review-uncertainty fields $U_S$ and $U_V$ are auxiliary outputs. The network $\mathcal{P}_\theta$ with parameters $\theta$ maps the input to six dense fields,
\begin{equation}
\mathcal{P}_{\theta}(X)=\{p_S,\,p_V,\,p_B,\,p_G,\,u_S,\,u_V\},\qquad
S\cap V\neq\varnothing\ \text{is admissible},\qquad
(S\cup V)\cap\Omega_s=\varnothing,
\label{eq:ontology}
\end{equation}
where $p_S$, $p_V$, $p_B$, and $p_G$ are the shock, vortex-core, background, and geometry probabilities and $u_S$ and $u_V$ are the shock and vortex review-uncertainty probabilities, all in $[0,1]$. The last condition in \cref{eq:ontology} excludes shock and vortex predictions from the exact body. The exact stereolithography (STL) surface of the body, or the solver wall mask, is applied after inference as a domain rule. It is never concatenated to the input, so any raw wall violation of the network remains measurable.

For clarity, the paper uses three related but distinct representations of a detected shock. The \emph{centerline} is a one-pixel-thick estimate of the shock locus and is used when a point-to-front distance is required. The \emph{envelope} is a finite-width, physically supported band grown around that centerline; it is used for overlap and stability measurements because a one-pixel line is very sensitive to subcell shifts. The \emph{boundary} is the one-pixel rim of the envelope and is used when the two sides of the detected band must be displayed or sampled. The envelope is an operational detection band, not a claim about the physical thickness of a shock. Likewise, a binary \emph{mask} is simply the set of raster pixels whose probability or physical-evidence score passes a stated threshold.

Several comparisons use \emph{weak references}. These are deterministic masks generated from compression, jump, rotation, topology and geometry rules; they are not hand-labelled ground truth. A Dice coefficient therefore measures agreement with the stated operational reference, with 1 denoting identical masks and 0 denoting no overlap. When object scores are reported, connected pixel groups are treated as objects and the component F1 score is the harmonic mean of one-to-one object precision and recall. These distinctions are important because pixel overlap, front localization and physical correctness are not interchangeable quantities.

A physics-only expansion proposal identifies regions with positive divergence, streamline-wise decreases in pressure and density, increasing speed, limited rotation and a fan-like topology anchored at a convex shoulder. Here \emph{pure expansion} means that these acceleration and pressure-drop signatures are present with little rotational evidence; \emph{mixed expansion/shear} denotes an expanding region that also carries appreciable shear or rotation. The expansion proposal is disjoint from the shock envelope by construction and accompanies the learned shock and vortex outputs as an annotation and audit class. The extended representation of \cref{sec:final-multistructure} adds independently learned wake/shear and expanding-flow outputs to this ontology.

\begin{figure}[!htbp]
\centering
\includegraphics[width=\textwidth]{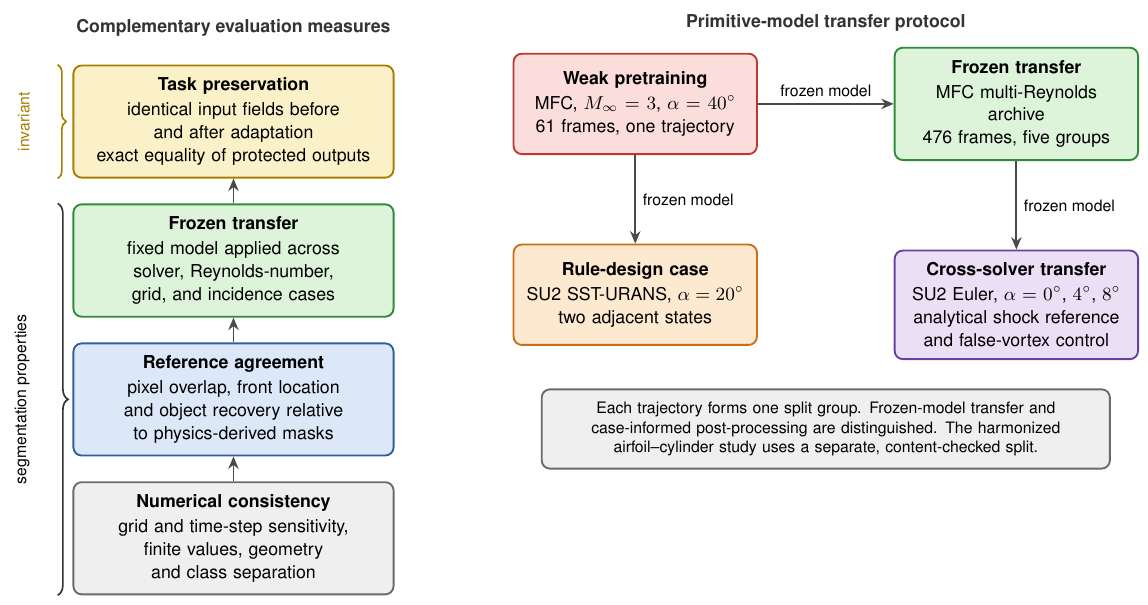}
\caption{Evaluation design and case organization. The left stack separates four questions that should not be conflated: agreement with physics-derived masks, stability to numerical sampling, transfer of a frozen model, and exact preservation of a protected output during adaptation. The right flowchart shows the primitive-model sequence: weak pretraining on the 40\dg{} MFC trajectory, a 20\dg{} SU2 rule-design case, frozen transfer to the multi-Reynolds archive, and cross-solver checks on low-incidence SU2 fields. Arrows indicate reuse of a frozen model rather than retraining. The harmonized airfoil--cylinder comparisons use the separate split and input definitions in \cref{sec:v4-study}.}
\label{fig:evidence}
\end{figure}

We evaluate five complementary quantities: consistency with physics-derived reference masks, localization relative to physical shock loci, stability under changes in numerical sampling, preservation of a frozen task during adaptation, and agreement with gas-dynamic references that do not depend on any learned or proposed mask, namely the analytical oblique-shock angles of the diamond airfoil and an empirical bow-shock correlation for the cylinder (\cref{sec:independent-gasdynamic}). \Cref{tab:evidence} lists the first four measures with their references, and \cref{fig:evidence} shows how they map onto the cases. The reference masks are constructed by the compression, jump and rotational criteria defined in \cref{sec:method}, so Dice and component scores measure agreement with those operational definitions. When a hybrid output reuses a physical proposal that also defines the reference, its score is identified as reference-conditioned agreement. Each evaluation set is characterized by what was withheld from optimization and by the nature of its reference. The development-test frames of the harmonized study are excluded from every optimizer update and scored against physics-derived references; the circle and ellipse trajectories are withheld in their entirety, with all thresholds fixed beforehand; and the analytical shock rays and the bow-shock correlation are independent of every learned or proposed mask. The flow fields come from the Multi-component Flow Code (MFC)~\cite{ref12} and from the SU2 suite~\cite{ref13}.

\begin{table}[!htbp]
\centering\small
\caption{Evaluation measures and their interpretation.}
\label{tab:evidence}
\begin{tabularx}{\textwidth}{l Y Y}
\toprule
Measure & Reference or comparison & Quantity assessed\\
\midrule
Reference agreement & Physics-derived masks and ignore regions & Pixel overlap and component recovery\\
Front localization & Native shock loci and compression envelopes & Position, coverage and finite-width representation\\
Numerical sensitivity & Matched grid, time-step and adjacent-state comparisons & Stability of the extracted structures\\
Task preservation & Identical inputs before and after branch-restricted adaptation & Exact equality of protected outputs\\
\bottomrule
\end{tabularx}
\end{table}
\FloatBarrier
\section{Flow cases, solvers, and leakage control}
\label{sec:data}
The study uses solver-native primitive fields from two independent CFD pipelines; \cref{tab:data-compact} summarizes the role of each evidence set. SU2 supplies diamond-airfoil fields at freestream Mach number $M_\infty=3$ and incidences $\alpha=0\dg$, $4\dg$, $8\dg$ and $20\dg$, including a circumferentially refined continuation used for the trailing-edge interaction audit. MFC supplies the viscous Mach-3 airfoil trajectory at $\alpha=40\dg$, the Reynolds-number sequence, and Mach-2.7 circular- and elliptical-cylinder trajectories. Lengths are referred to the airfoil chord $c$ or the cylinder diameter $D$, the freestream speed is $U_\infty$ with freestream density $\rho_\infty$ and pressure $p_\infty$, and $\Rey_c$ denotes the chord-based Reynolds number. Grid labels state the spatial resolution: f180 and f270 mean 180 and 270 cells per airfoil chord, while f90 and f180 mean 90 and 180 cells per cylinder diameter. Cylinder runs are further identified by their Courant--Friedrichs--Lewy (CFL) number, which controls the explicit time-step size. The stored time $t$ is the nondimensional solver time of each case, recorded with every state.

Throughout the paper, \emph{native grid} means the mesh on which the CFD solver produced the field, whereas \emph{raster} means the regular Cartesian image array on which the neural network and image-space diagnostics are evaluated. The two should not be confused. In particular, the core 40\dg{} MFC trajectory is converted to a $900\times990$ observation raster by taking every third native-grid cell, reducing the sampling from 270 to 90 cells per chord. We call this the \emph{stride-3 raster}: ``stride 3'' refers only to this every-third-cell sampling operation, not to a convolutional-network stride. Other case families use the raster sizes stated with their results. The exact wall/solid mask stays outside the neural input and is applied only as a domain constraint.

For visual comparison we use \emph{numerical schlieren}, a deterministic display of density-gradient magnitude rather than an experimental optical schlieren image. When a caption says \emph{fixed-scale schlieren}, the same density-gradient transfer function and contrast limits are used for all panels being compared; the display is not rescaled separately to make a weak feature look stronger.

Training, validation and test membership is grouped by complete trajectory and geometry family, so adjacent snapshots from one trajectory never cross a split boundary. The 84 airfoil and 25 cylinder training fields form the weak-label training bank; duplicate initial fields are excluded from the 43-field validation and 43-field development-test sets. The circle, ellipse and changed-incidence trajectories used for prospective transfer were computed after the models and their thresholds had been fixed, and none of their results was inspected before evaluation. The continued SU2 fields and the analytical gas-dynamic references are never used for optimizer updates. Solver equations, meshes, numerical fluxes, time integration, nondimensionalization, state conversion and the complete split manifests are given in Supplementary Section S1.

\begin{table}[!htbp]
\centering\small
\caption{Roles of the principal evidence sets. A ``test'' field is withheld from optimizer updates; only the prospective whole-case and analytical rows are also withheld from configuration inspection.}
\label{tab:data-compact}
\begin{tabularx}{\textwidth}{>{\raggedright\arraybackslash}p{0.17\textwidth} >{\raggedright\arraybackslash}p{0.22\textwidth} >{\raggedright\arraybackslash}p{0.25\textwidth} Y}
\toprule
Set & Geometries/solver & Role & Reference type\\
\midrule
Harmonized MFC & airfoil and cylinder & grouped train/validation/test & stored weak masks\\
Prospective transfer & circle, ellipse, incidence and Reynolds changes & whole-trajectory evaluation with fixed models & audits fixed in advance\\
Continued SU2 & diamond airfoil, $0$--$20\dg$ & cross-solver front/fan audit & analytical rays and field diagnostics\\
Analytical controls & ideal shocks, vortices and fans & independent or generator-bounded controls & closed-form relations\\
\bottomrule
\end{tabularx}
\end{table}
\FloatBarrier
\section{Physics-audited multi-task segmentation}
\label{sec:method}
\subsection{Outputs and branch-specialized network}
The primary network is a compact three-level U-Net-like encoder with separate, complete decoders for shocks and vortex cores (\cref{fig:arch}). Independent sigmoid heads produce probabilities for shock, vortex core, background, geometry and two review-uncertainty fields. Because shock and vortex use separate sigmoids rather than a mutually exclusive softmax, the same pixel may legitimately belong to both classes.

Three learned configurations appear repeatedly in the results. The \emph{primitive encoder} (PE) uses only the four fields in \cref{eq:input}. The \emph{branch-isolated refinement} (BIR) augments PE with rotational diagnostics, but feeds them only into the vortex decoder. The added residual adapters are initialized to zero, so BIR starts with exactly the PE response and then learns only the allowed correction; the shock pathway is unchanged. PE and BIR support the single-trajectory transfer experiments. The \emph{harmonized joint} model (HJ) uses seven freestream-referenced channels in a shared encoder: clipped log-density and log-pressure, the velocity components parallel and perpendicular to the freestream, nondimensional vorticity, the deviatoric $Q$ criterion, and swirling strength. The last three encode local rotation and strain; the exact nondimensional scaling and clipping constants are given in Supplementary Section S2. HJ supports the controlled airfoil--cylinder comparisons. HJ has 268,496 parameters in the matched study; the conventional multi-label U-Net baseline has 267,639. The central adaptation rule is structural: when the shock task is refined, every parameter and intermediate feature on which the protected vortex output depends remains fixed. We verify the resulting equality directly on every protected field rather than inferring preservation from an average score.

\begin{landscapefigure}
\includegraphics[width=\linewidth,height=\lsfigureheight,keepaspectratio]{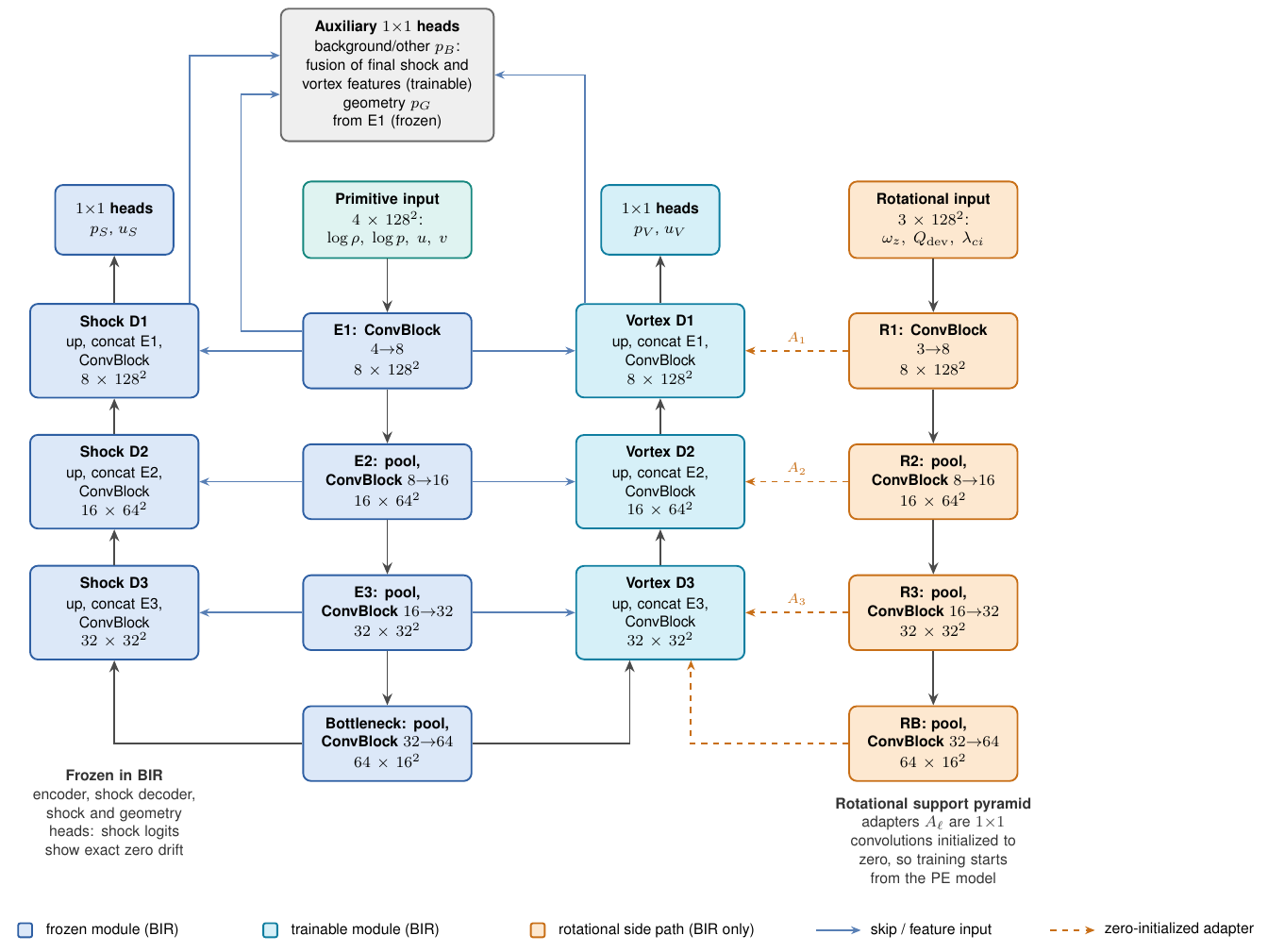}
\caption{Branch-specialized architecture used for PE and BIR. The central blue path encodes the four primitive inputs $(\log\rho,\log p,u,v)$ at progressively coarser spatial scales. The left decoder produces shock-related outputs; the right decoder produces vortex-related outputs. Orange blocks form the rotational side path used only by BIR, and the dashed orange adapters $A_1$--$A_3$ inject those rotational features into the vortex decoder. Blue modules marked as frozen retain the PE weights, whereas cyan modules are trainable during BIR refinement. Thus the rotational refinement can change the vortex branch without changing the shock branch. The auxiliary $1\times1$ heads convert decoder features to the six probability fields defined in \cref{eq:ontology}.}
\label{fig:arch}
\end{landscapefigure}

\subsection{Physics evidence and class separation}
\label{sec:method-rh}\label{sec:method-shockproposal}\label{sec:method-vortex}\label{sec:method-expansion}
Shock proposals combine several pieces of evidence rather than treating every large gradient as a shock. Negative velocity divergence identifies local compression; aligned jumps in pressure and density test whether the thermodynamic change has the expected sign; and shock-normal non-maximum suppression keeps only the local ridge of a broad numerical front instead of every high-gradient pixel. Candidates supported only by the wall or image boundary are rejected. Local Rankine--Hugoniot residuals then measure how closely the sampled upstream and downstream states satisfy the conservation jump relations. We use that residual as supporting evidence rather than as a universal hard gate, because near-wall and interacting fronts do not always provide clean two-sided samples. Entropy is also supporting evidence only: a downstream entropy layer or wake can remain strong after the actual shock surface has passed.

Vortex proposals similarly combine diagnostics with different physical meanings. Vorticity measures local rotation, swirling strength tests for locally complex eigenvalues associated with swirling motion, the deviatoric $Q$ criterion compares rotation with shear-producing strain after removing isotropic compression, and the circulation-based test measures angular coherence around a neighborhood. Spatial coherence and temporal persistence are then used to reject isolated noisy pixels. Strong near-wall vorticity is treated as shear unless the independent rotation tests support a compact core, and a candidate is not rejected merely because it overlaps a shock. Shock, core and background form the primary segmentation classes; wake/shear is evaluated as a separate auxiliary output.

The expansion pathway first predicts an expanding-flow response. A geometric fan decoder then proposes limiting rays from connected responses and wall corners. Its initial component path checked for local supersonic flow but omitted a characteristic test, and its additional-origin path used a 12-degree tolerance; the high-incidence analysis in \cref{sec:results} identifies the resulting false rays. A 3-degree local-angle check is reported as a necessary diagnostic, and analytical rays are always displayed explicitly as references. A broad region that accelerates after a curved bow shock therefore stays an expanding-flow region and is not classified as a centred fan. The network and optimization equations are given in Supplementary Section S2; the fan-decoder implementation and its physical audit are documented in Supplementary Section S3.

\subsection{Controlled comparisons and evaluation}
\label{sec:v4-study}\label{sec:shock-adaptation-method}
All architecture comparisons use identical case splits, patch banks, optimization steps and the same three initialization seeds. Five learned variants are compared: the harmonized joint network; the same network trained without projected conflicting gradients (PCGrad), which normally removes the component of one task gradient that directly opposes the other in the shared encoder; a primitive-channel-only variant; the SegFormer-B0 transformer \cite{v4Xie2021SegFormer} adapted to seven inputs and six logits; and a temporal-target ablation in which frozen proposals from a temporal-association teacher supply training targets. A capacity-matched conventional U-Net is added in \cref{sec:capacity-unet}.

For every model we keep three products conceptually separate. The \emph{machine-learning-only} (ML-only) mask is obtained by thresholding the learned probability at an operating threshold $\tau$ selected only on validation fields. The \emph{physics-only} mask is generated by the fixed diagnostic rules without neural probabilities. The \emph{hybrid} starts from the ML-only mask and adds only pixels that satisfy the stated physical-support rule. This separation makes it possible to see whether an improvement comes from the network, the physical proposal or their combination. Pixel Dice measures overlap of two masks; component precision and recall measure one-to-one recovery of connected objects, and their harmonic mean is the component F1 score. We report those quantities alongside front localization, finite-width envelope overlap, exact protected-task equality, temporal continuity and full-field overlays. Thresholds are never adjusted on a test frame, and missing contours are never completed manually.

Shock adaptation is the task-preserving refinement of the HJ model. The shared encoder and the complete vortex pathway are frozen, and only the shock decoder and its final projection are updated, so the full-field vortex probability of the adapted model equals that of the base model by construction. Two configurations are retained. The clean-target configuration applies 600 additional updates on 2,400 training patches with an expanded full-front airfoil teacher, unchanged cylinder targets, geometry negatives and boundary and ambiguity exclusions. The noise-augmented configuration starts from those weights and applies 300 further updates to 1,200 patches, 414 of them clean and 786 with velocity perturbations of 0.5, 1 or 2\% of $U_\infty$; the differential channels are recomputed from the perturbed velocity while density, pressure and the clean reference labels stay fixed. The shock operating point is selected on validation data at 0.97 and the protected vortex operating point stays at 0.85. Because this experiment changes the supervision and the optimization budget together, it is reported separately from the five-model matched-budget comparison, and both the base and the adapted models are compared against the same expanded reference when front recovery is assessed. The complete specification is given in Supplementary Section S2.
\section{Numerical results}
\label{sec:results}
The transfer experiments cover weak attached shocks, separated airfoil flows and a detached cylinder shock. The BIR model, its normalization statistics and its thresholds were fixed on the 40\dg{} MFC trajectory before any other case was processed, so the SU2 and multi-Reynolds results below are applications of one trained model. The single post-processing rule that was set on another case, the image-boundary rule of \cref{sec:results-su2urans}, is identified where it is used. We first examine primitive-encoder transfer to the 20\dg{} SU2 case, the separated 40\dg{} airfoil and the cylinder. We then present the controlled harmonized comparisons, independent shock-reference checks and extended wake and expansion outputs. Further development and transfer results are provided in the Supplementary material.

\subsection{Unsteady SU2 SST-URANS solution at 20\dg{} incidence}
\label{sec:results-su2urans}
The $20\dg$ case is an unsteady Reynolds-averaged Navier--Stokes solution with the shear-stress-transport turbulence model (SST-URANS). It combines a lower leading-edge shock, an upper leading-edge expansion, shoulder expansions and near-wall rotation. With the $8\dg$ diamond half-angle, the upper leading corner turns the incident flow away from the surface by $12\dg$, so ideal shock--expansion theory places an expansion fan, not a shock, at the upper leading edge. The computed field contains one feature that the inviscid construction does not: a weak compression ahead of that expansion, about 19 percent above the sampled freestream pressure. Its angular position is the same in the two stored states. Possible origins include viscous interaction at the tip, mesh-scale tip geometry and residual unsteadiness (Supplementary Section S3). The neural-seeded hybrid marks this compression and the frozen LocalFront prediction does not. Ideal turning theory alone is insufficient to determine which response is appropriate, so both are reported. LocalFront is a separate 9,026-parameter convolutional network with finite spatial context that predicts front and transition-band probabilities; it was trained for 600 updates, with validation-selected thresholds of 0.8 and 0.9.

The original leading-edge attachment rule removes the 4,476 and 4,475 interpolation-boundary pixels of the two stored states, but it also suppresses trailing compression branches. Removing it admits an upper trailing branch while leaving the leading-edge compression response in place. The three-panel whole-field comparison of the schlieren, the neural-seeded hybrid without the veto and the frozen LocalFront prediction is a development diagnostic and is given in Supplementary Section S5; the same field is compared with the ideal shock--expansion construction in \cref{fig:fan-alpha20}. The adjacent-state Dice of 0.99285 measures the repeatability of the selected masks between the two states. The physics expansion proposal has 2,199 and 2,191 pixels, and BIR identifies a repeatable seven-pixel near-wall rotational candidate near $(x/c,y/c)=(0.990,0.016)$. The full-field maps provide the spatial context needed to interpret these counts.

\subsection{Core MFC trajectory: classical baselines}
\label{sec:results-mfc-baselines}
The separated 40\dg{} trajectory is the most demanding case and the one on which the model was developed. We first ask what the fixed Fujimoto sensor does on this flow before any learned model or connectivity repair is introduced. On the stride-3 observation raster defined in \cref{sec:data}, the reproduced sensor returns a median of 2,639 connected shock components per noninitial frame and a Dice coefficient of 0.0760 against the weak shock mask over the 61-frame trajectory (\cref{fig:baselines}). Here a connected component is an eight-connected group of neighboring detected pixels; thousands of components therefore indicate a highly fragmented response rather than thousands of physical shocks.

To determine whether this fragmentation is created by the factor-three downsampling, we rerun the sensor in a native-grid region of interest at 270 cells per chord (\cref{fig:fujimoto-native}). At the three sampled times the native grid returns 2,109, 3,023 and 3,516 components, and the largest connected piece contains only 1.8 to 3.3\,\% of all detected cells. The stride-3 raster returns 917, 1,498 and 1,566 components at the same times. Thus the finer grid reveals more pressure-edge candidates, but it still does not assemble them into one coherent bow-shock front; responses remain distributed across the wake, wall-adjacent gradients and numerical wave trains. We intentionally show the raw sensor output without component-length filtering, because filtering would hide the behavior being evaluated. The comparison therefore diagnoses fragmentation under the sensor's own operating principle rather than providing an equal-post-processing benchmark.

The separate compression/jump proposal of \cref{sec:method-rh}, after shock-normal non-maximum suppression, has 36.24\,\% local Rankine--Hugoniot support according to the Fujimoto consistency test. The topology-and-time vortex baseline produces 63 persistent tracks, of which 47 overlap the shock band at some time. Its weak Dice of 0.8656 should be interpreted cautiously because the proposal and weak vortex reference share rotational criteria.

\Cref{fig:baselines} is deliberately plotted without hiding these failure modes. The scattered red marks in the Fujimoto column are not separate physical shocks: they are disconnected local detections triggered by strong pressure edges in the wake, near the wall and in numerical wave trains. In the physics-proposal column, gold does not denote another flow structure; it marks only those red shock-candidate pixels that also satisfy the local Rankine--Hugoniot consistency test. The yellow--orange marks in the rightmost agreement column have a third meaning: they are pixels selected by the physics proposal but not by the neural mask. Blue pixels are the converse, and green pixels are selected by both. Thus isolated red or orange/gold points in this figure are evidence of sensor disagreement or local physical support, not additional shocks or vortices. The middle row is a difficult, quality-control-flagged state whose weak shock label was excluded from the shock loss; it is shown because it exposes the behavior of the methods under an adverse field rather than because it defines a new physical regime.

A fragmented cellwise shock flag can still be useful for flux switching and local post-processing. Continuous front segmentation and temporal object tracking require additional spatial information, particularly when shocks and vortex cores overlap. This difference in purpose motivates the learned spatial representation used here, with conservation-based sensing retained as physical evidence.

\begin{landscapefigure}
\includegraphics[width=\linewidth,height=\lsfigureheight,keepaspectratio]{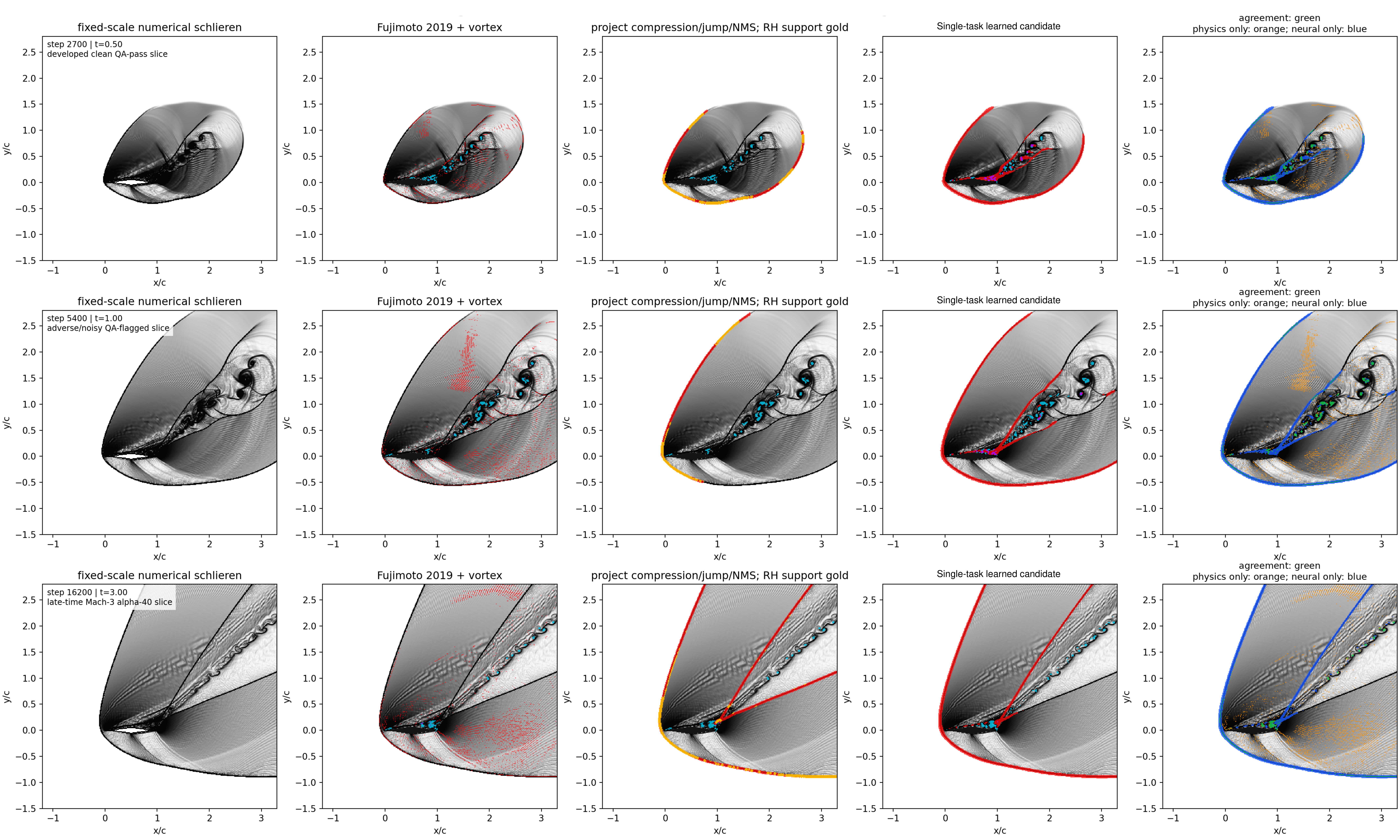}
\caption{Classical shock sensing, physical proposals and a shock-only learned baseline on three states of the core MFC trajectory: a relatively clean state (top), a quality-control-flagged adverse state whose weak shock label was excluded from the shock loss (middle), and a late separated state (bottom). Column 1 is fixed-scale numerical schlieren, i.e. the same density-gradient visualization for all rows. Column 2 overlays the raw Fujimoto pressure-edge shock-cell detections in red and the independent vortex proposal in cyan. The dispersed red marks are fragmented sensor responses to strong non-shock gradients in the wake, near the wall and in numerical wave trains; they do not represent many physical shocks. Column 3 shows the paper's physics-only shock proposal after compression, pressure/density-jump and shock-normal non-maximum suppression, which keeps the local ridge of a broad numerical front. Gold is not a second structure: it is the subset of the red proposal whose sampled two-sided states also satisfy the local conservation-jump (Rankine--Hugoniot) check. Column 4 is the single-task, shock-only neural baseline trained on the same core trajectory. Column 5 compares the physics proposal with that neural mask pixel by pixel: green means both select the pixel, orange/yellow means physics only and blue means neural only. Scattered orange/yellow marks therefore visualize disagreement, not another flow feature. The figure is a diagnostic of continuity and failure modes; neither proposal is treated as ground truth.}
\label{fig:baselines}
\end{landscapefigure}

\begin{landscapefigure}
\includegraphics[width=\linewidth,height=\lsfigureheight,keepaspectratio]{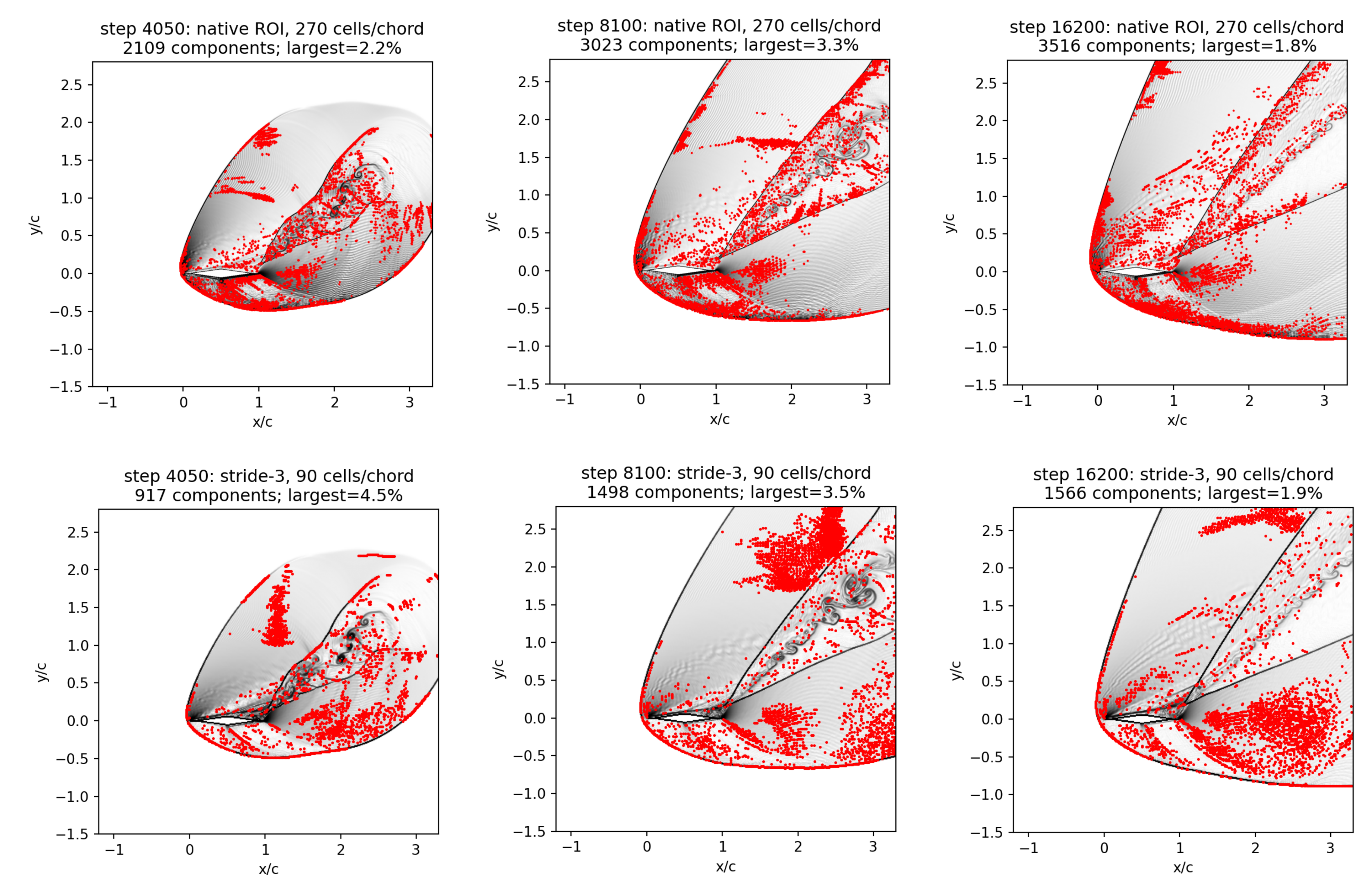}
\caption{Resolution audit of the Fujimoto shock sensor. The top row uses the native MFC grid in the displayed region of interest, with 270 cells per chord. The bottom row uses the stride-3 raster, obtained by retaining every third native-grid cell and therefore sampling the same flow at 90 cells per chord. Columns correspond to $t=0.75$, $1.5$ and $3$ (solver steps 4050, 8100 and 16200, as labelled). Red pixels are cells classified as shock by the sensor. The number above each panel is the count of eight-connected detected components, and the accompanying percentage is the fraction of all detected pixels contained in the largest component. A coherent shock would place a substantial fraction of its pixels in a small number of long components; here the largest share remains only a few percent. Because fragmentation is already present on the native grid, it cannot be attributed to stride-3 downsampling alone.}
\label{fig:fujimoto-native}
\end{landscapefigure}

\subsection{Reynolds-number and grid transfer}
\label{sec:results-multire}
We apply the fixed BIR model and the shock-object workflow to every field of the 476-frame archive (\cref{tab:multire}, \cref{fig:re-outputs}). For $t\ge1$ the upper and lower regional long fronts are present in 100\,\% of the eligible frames of all five case groups. The median adjacent-frame Dice of the shock envelope is 0.884, 0.876, 0.806, 0.794 and 0.896 for $\Rey_c=10^4$ (f180), $10^4$ (f270), $5\times10^4$, $10^5$ and the retained $10^6$ sequence. The median number of vortex-positive pixels rises from 116 and 108 on the two $\Rey_c=10^4$ grids to 650, 672 and 687 at the three higher Reynolds numbers. This roughly sixfold increase matches the visibly richer separated wake in the complete archive. Because the f270 grid and the $\Rey_c=10^6$ time window differ from the other cases, we read the trend qualitatively. The detector finds lower mixed expansion and shear evidence in every eligible frame and upper mixed evidence in 52.5 to 100\,\% of the frames depending on the case, whereas the pure-expansion prevalence is lower and non-monotone.

\begin{table}[!htbp]
\centering
\caption{Frozen transfer over the complete multi-Reynolds archive for $t\ge1$. Temporal Dice is the median Dice coefficient of the shock envelope between adjacent saved states; the vortex column is the median number of vortex-positive pixels per frame on the $512\times512$ raster; the last column gives the share of eligible frames with mixed expansion and shear evidence in the upper and lower regions.}
\label{tab:multire}
\small
\setlength{\tabcolsep}{4pt}
\begin{tabularx}{\textwidth}{p{0.24\textwidth} Z Z Z Z Z}
\toprule
Case group & Eligible frames & Upper / lower long front present & Median temporal Dice & Median vortex pixels & Upper / lower mixed expansion \\
\midrule
$\Rey_c=10^4$, f180 & 101 & 100\,\% / 100\,\% & 0.884 & 116 & 61.4\,\% / 100\,\% \\
$\Rey_c=10^4$, f270 & 101 & 100\,\% / 100\,\% & 0.876 & 108 & 52.5\,\% / 100\,\% \\
$\Rey_c=5\times10^4$, f180 & 51 & 100\,\% / 100\,\% & 0.806 & 650 & 94.1\,\% / 100\,\% \\
$\Rey_c=10^5$, f180 & 51 & 100\,\% / 100\,\% & 0.794 & 672 & 98.0\,\% / 100\,\% \\
$\Rey_c=10^6$, f270 (retained) & 112 & 100\,\% / 100\,\% & 0.896 & 687 & 100\,\% / 100\,\% \\
\bottomrule
\end{tabularx}
\end{table}

The regional-presence test requires at least 50 accepted centerline pixels in each of the upper and lower far-front regions. Every eligible frame meets this criterion. The complementary envelope-overlap statistic then measures how much the finite-width front changes between adjacent saved states. Its median stays between 0.794 and 0.896, whereas the thin centerline is more sensitive to phase and raster alignment. The matched $\Rey_c=10^4$ grid pair quantifies this difference over 101 paired times (\cref{tab:gridcontrol}): the median paired Dice is 0.810 for the finite-width envelope but only 0.557 for the one-pixel centerline, with 0.738 for the vortex mask and 0.590 for the mixed-expansion mask. The envelope tolerates raster displacement better. Higher overlap under dilation does not, however, establish more accurate localization or numerical convergence, and the maximum half-width of $0.045c$ allows a full width of $0.09c$.

\begin{table}[!htbp]
\centering
\caption{Matched $\Rey_c=10^4$ grid-control audit over 101 paired times on the common $512\times512$ raster. Each entry is the Dice coefficient between the f180 and f270 masks of the same structure at the same time; the range gives the 5th and 95th temporal percentiles computed from the archived numerical distribution.}
\label{tab:gridcontrol}
\small
\begin{tabular}{l c c}
\toprule
Structure & Median paired Dice & 5th to 95th percentile \\
\midrule
Shock centerline (one pixel) & 0.5569 & 0.5142 to 0.6051 \\
Shock envelope (finite width) & 0.8104 & 0.7979 to 0.8259 \\
Vortex mask (BIR) & 0.7381 & 0.6126 to 0.8099 \\
Mixed expansion mask & 0.5902 & 0.5084 to 0.7656 \\
\bottomrule
\end{tabular}
\end{table}

\begin{landscapefigure}
\includegraphics[width=\linewidth,height=\lsfigureheight,keepaspectratio]{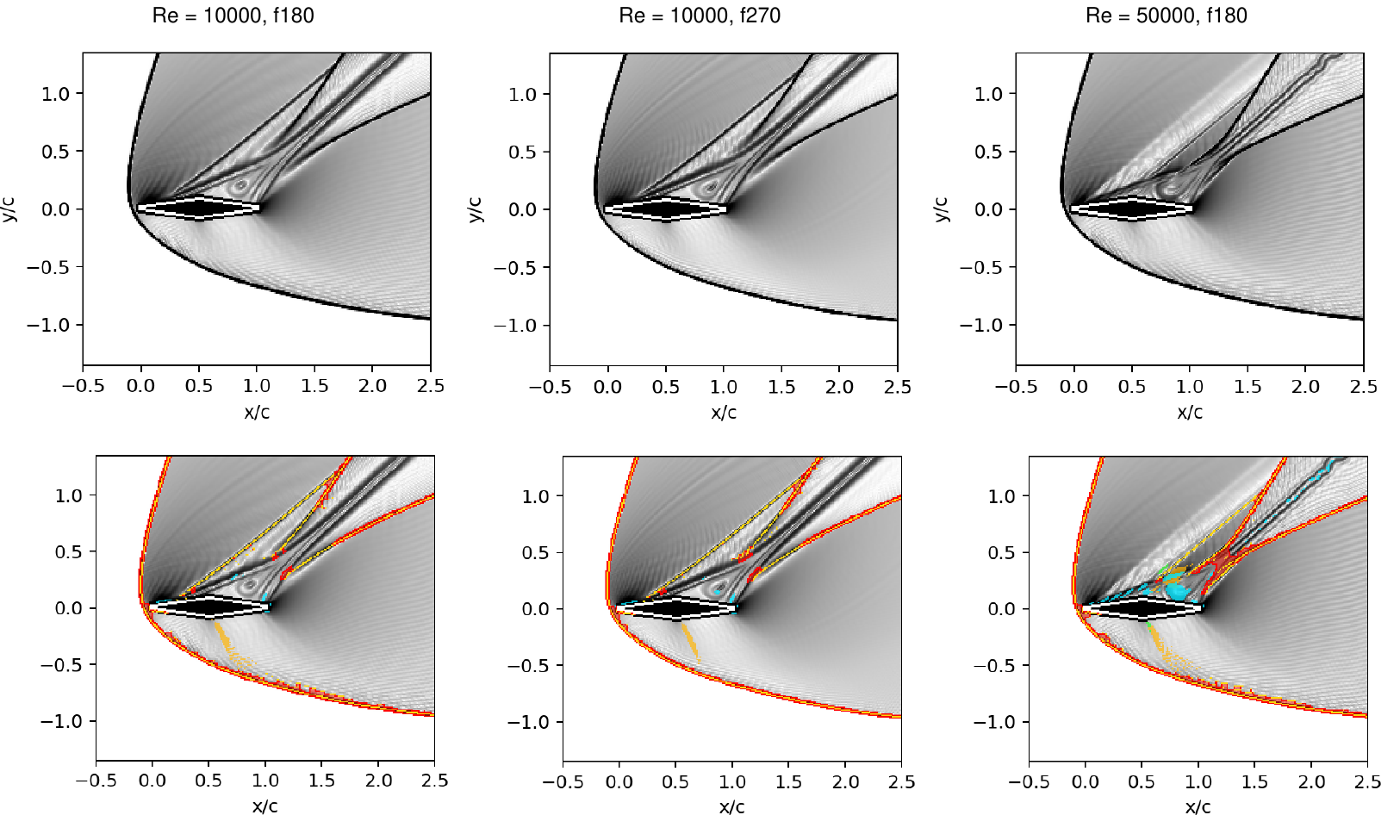}
\caption{Frozen branch-isolated-refinement (BIR) transfer across Reynolds number and grid resolution. Columns show $\Rey_c=10^4$ on f180 (180 cells/chord), the matched $\Rey_c=10^4$ case on f270 (270 cells/chord), and $\Rey_c=5\times10^4$ on f180. The top row is fixed-scale numerical schlieren. In the bottom row, red is the finite-width shock envelope and its rim, and yellow is the one-pixel centerline extracted from that same envelope for localization; yellow is therefore not a second shock class. Cyan is the BIR vortex-core mask, green is a pure-expansion proposal with little rotational support, and gold is an expanding region that also carries appreciable shear/rotation. The larger cyan response in the higher-Reynolds-number case follows the visibly richer separated wake, while green/gold patches away from the dominant red fronts are expansion/shear responses rather than additional shocks. The full displayed field is retained so that these wake responses and false positives remain visible.}
\label{fig:re-outputs}
\end{landscapefigure}
\begin{landscapefigure}[\ContinuedFloat\captionsetup{labelformat=continued}]
\includegraphics[width=\linewidth,height=\lsfigureheight,keepaspectratio]{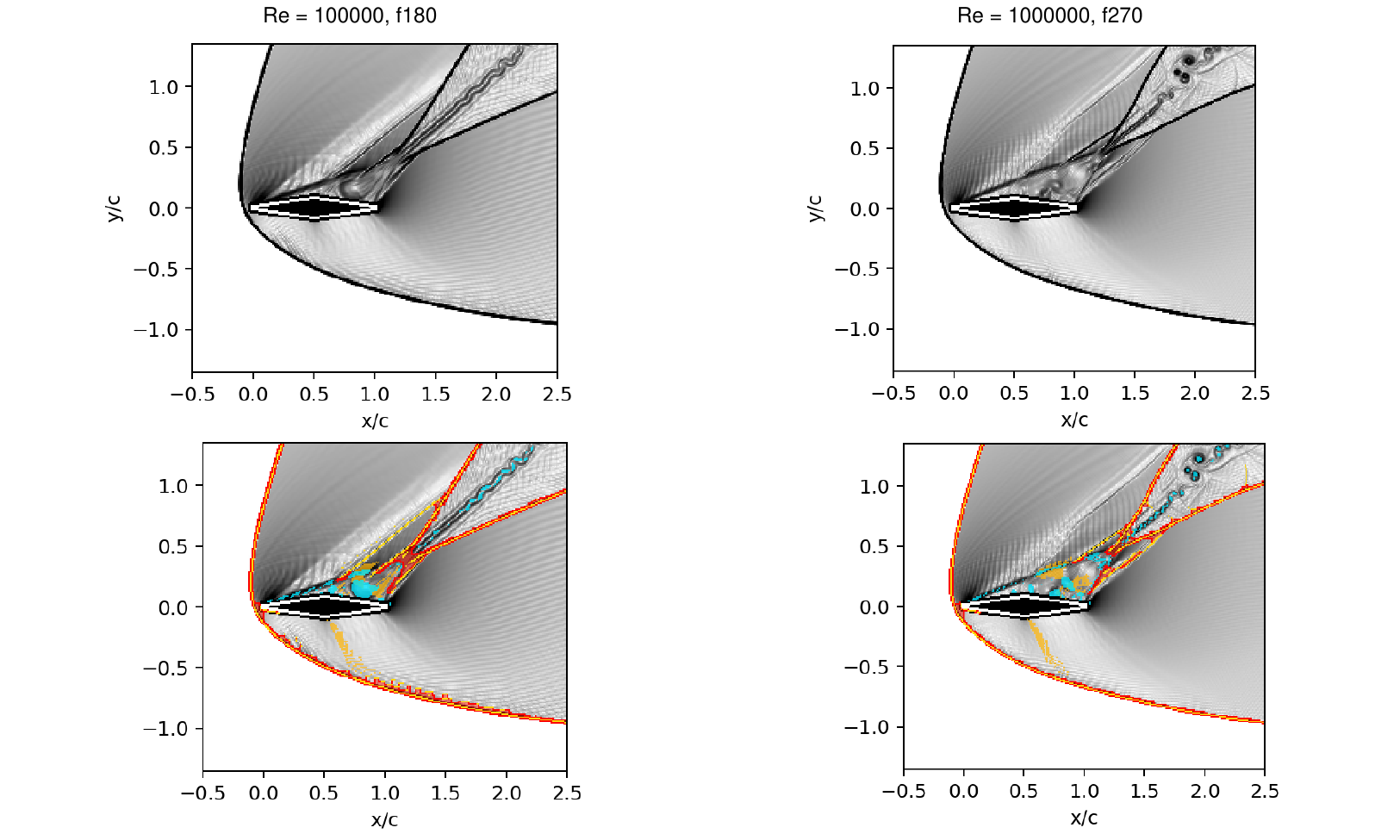}
\caption[]{Continuation of the frozen BIR transfer in \cref{fig:re-outputs}, here for $\Rey_c=10^5$ on f180 and the retained $\Rey_c=10^6$ sequence on f270. Colors and objects have the same meanings as in the first part: red shock envelope/boundary, yellow centerline derived from that envelope, cyan vortex-core mask, green pure expansion and gold mixed expansion/shear. Yellow therefore traces the location assigned to the same red shock band rather than a separate structure. The top row again uses the same numerical-schlieren scale, and the full field is shown without cropping away difficult wake regions, so isolated auxiliary responses remain visible.}
\end{landscapefigure}

\label{sec:results-qa}
All 476 archive frames pass the structural checks after the final shock-envelope construction: zero overlap of the shock, vortex and expansion masks with the exact geometry, zero shock and expansion overlap, every centerline pixel inside its envelope, finite probabilities everywhere, and a background field that is the exact complement of the foreground classes. The SU2 outputs satisfy the same rules. The MFC core trajectory, the multi-Reynolds archive and each SU2 family are treated as distinct case groups throughout.

\subsection{Cylinder transfer: localization versus wake interpretation}
\label{sec:results-cylinder}
The PE and BIR models are transferred to all three cylinder trajectories without cylinder training or recalibration; \cref{tab:cylinder-transfer} lists the outcome. Median ML-only coverage of the physical bow-shock envelope is 0.9378--0.9481, whereas median precision against that proposal is only 0.4933--0.5015. The lower precision reflects responses beyond the bow front, including the downstream recompression fronts and wake structures that lie outside the bow-envelope reference; these are examined in \cref{sec:final-multistructure}. An anchored, neural-seeded hybrid keeps the physical bow envelope in all 30, 80 and 80 post-initialization states.

\begin{table}[!htbp]
\centering\small
\caption{Frozen-model transfer to the three circular-cylinder trajectories. Dice is mask overlap with the entropy-free physics bow-envelope proposal; coverage is the fraction of that reference envelope reached by the prediction; precision is the fraction of predicted pixels lying inside the reference. Each is the median over saved times after initialization. Stand-off is the final distance from the cylinder nose to the detected bow-shock front, normalized by the diameter $D$.}
\label{tab:cylinder-transfer}
\begin{tabular}{lrrrr}
\toprule
Run & Dice & Coverage & Precision & Stand-off/$D$\\
\midrule
f90, CFL 0.20 & 0.6514 & 0.9415 & 0.4934 & 0.40556\\
f90, CFL 0.10 & 0.6521 & 0.9378 & 0.4933 & 0.40556\\
f180, CFL 0.20 & 0.6598 & 0.9481 & 0.5015 & 0.39444\\
\bottomrule
\end{tabular}
\end{table}

The nose of the bow shock settles over the recorded interval. At the final state the sampled pressure and density jumps agree with the normal-shock relations to within 1.69\% and 2.32\%, and the nose pressure coefficient to within 0.70\%, while the sampled downstream Mach number lies 37.16\% below the normal-shock value, consistent with sampling inside the decelerating stagnation region rather than immediately behind the front. The wake, by contrast, has not reached a periodic state within the recorded interval: the three runs do not share a dominant shedding frequency, and the symmetric persistent-track match fractions are 0.1500 between the two time steps and 0.0741 between the two grids at matched step. We therefore interpret the extracted wake trajectories at the recorded numerical resolution. The segmentation target is the set of structures present in the supplied numerical field, with its solver and resolution recorded, so the HJ model and the shock adaptation keep the wake candidates rather than enforcing an empty vortex mask on the basis of the Euler equations alone.
\input{v4_results}
\input{corrected_control_extension}
\input{independent_gasdynamic_validation}
\input{compact_transfer_atlas}
\input{compact_multistructure}
\section{Discussion}
\label{sec:discussion}
The whole-field comparison of the $20\dg$ case (\cref{sec:results-su2urans}; Supplementary Section S5) illustrates a limitation of geometric filtering. Requiring attachment to a prescribed origin can remove a trailing compression front despite neural seeds and jump evidence. Relaxing that rule recovers the component, but does not resolve every ambiguity: the hybrid retains the leading-edge compression response, and the frozen learned alternative adds responses near the wall. Completeness, precision and task preservation must therefore be assessed separately.

Relative to the physics-only sensor of Fujimoto et al.\ \cite{ref1}, the framework represents compressible-flow structures as learned, overlapping classes and objects in which shock and vortex identities coexist and may overlap, while the classical compression and Rankine--Hugoniot criteria stay visible as evidence. Relative to image-level detection studies \cite{ref5,ref6,ref22}, the outputs are spatially resolved and aware of the solver field, so front length, boundary thickness, object identity, wall proximity and track persistence can be evaluated directly.

The most specific contribution is branch-isolated physics support and, more generally, task-preserving adaptation. A shared primitive representation is kept, but rotational diagnostics are injected through zero-initialized residual adapters into the vortex decoder only, and shock supervision is refined through the shock decoder only. Zero output drift confirms preservation of the frozen computational path, an established freezing principle. It does not establish accuracy, but demonstrates that the vortex refinement in BIR and the front recovery during shock adaptation leave the other task unchanged.

The finite-width shock envelope is a second practical contribution. On the paired $\Rey_c=10^4$ grids it is markedly more stable than the one-pixel ridge (median paired Dice 0.8104 against 0.5569), while the ridge is kept for point-distance evaluation. The interference audit documents conflicting gradients in 27 to 53\,\% of the audited steps. The matched no-PCGrad comparison of \cref{sec:v4-results} then shows that projecting them has geometry-dependent rather than uniform effects on the segmentation scores, which is why we combine gradient projection with the dependency restriction rather than relying on it alone.

The transfer experiments show why a single scalar score is inadequate for this problem. Across the 476-frame archive both long-front regions contain at least 50 accepted pixels in every eligible frame, which is a regional-presence test rather than full-front coverage; the shock-envelope stability stays high; and the vortex mask grows by roughly a factor of six from $\Rey_c=10^4$ to the higher Reynolds numbers while expansion and shear evidence become increasingly mixed. The separate outputs retain these distinctions, while the field maps reveal failure modes that scalar metrics alone would obscure. The fixed pressure-edge and Rankine--Hugoniot baseline produces disconnected responses in the complex wake even at native resolution. A primitive-only joint vortex head underfits the diagnostic proposal, motivating the branch-specific introduction of physical diagnostics. Frozen neural thresholds can activate on interpolation and domain boundaries; component anchoring together with interior physics evidence removes these components. Soft support increases the number of near-wall positives, which contain coherent rotation or shear and are separated through local structure and temporal evidence. A directly transferred expansion rule accepts rotational wake regions, so cross-solver thresholds have to be re-derived for each solver. Finally, the freestream orientation drives the standardized $v$ channel far outside the training support at low incidence, which is what led to the freestream-aligned inputs of the harmonized study. On the continued SU2 fields those inputs, together with shock adaptation, give a mean normal displacement of about 0.003 chord on the five finite ray segments over $0.10\leq x/c\leq0.40$, not over complete fronts.

The workflow contains many physical quantities, but each has a bounded role. Compression and jumps generate weak shock seeds and audit front plausibility; rotation and topology generate vortex review seeds; exact geometry excludes impossible positives; and the expansion logic defines a separate candidate class. The learned shock and vortex probabilities remain the primary dense decisions, the soft confidence is always reported beside the raw learned output, and the physics-only baselines are stored independently. This separation shows which spatial structures come from the learned representation, which are supplied by classical proposals and which are added by the hybrid. A purely threshold-based detector would inherit the brittleness of the fragmented classical sensor and of the transferred expansion rule, whereas an unconstrained image network can activate on airfoil or interpolation boundaries and may sacrifice one task to improve the other. The proposed representation combines measured gradient control, branch-specific physical information and explicit post-inference checks to address these difficulties.

The compact architecture makes full-field iteration possible without a large accelerator. PE contains 171,926 trainable parameters and completed its $6\times60$ training run in 665~s on a central processing unit (CPU); BIR contains 251,334 parameters, updates 128,187 of them and needed 356~s. Full-domain inference is tiled, so memory scales with the $384^2$ tile and is independent of the $900\times990$ raster or the native solver field. Every output is checked automatically for overlap with the body, invalid probabilities, overlap between shock and expansion masks and centerline pixels outside their envelope; the archives and configuration records that support exact replay are described in the Data and code availability statement.

The references and flow cases define the scope of the conclusions. The quantitative segmentation comparisons use physics-derived weak references, and are interpreted alongside full-field overlays, localization measures and the two label-independent references. Three seeds characterize initialization variability on fixed cases. The two geometries and two solvers cover attached and detached shocks, shoulder expansions, separated shear layers and inviscid wakes, and the withheld circle and ellipse trajectories extend the transfer to withheld parameter combinations within these geometry families. Deployment on a new geometry requires the same primitive variables, reference scales, coordinates and geometry masks, and the raster-based component cutoff and dilation widths scale with the resolution.

The gas-dynamic references constrain shock geometry. On the continued SU2 fields the native shock angles agree with theory to within $0.42\dg$, and the frozen shock-adapted models cover the five tested ray segments over $0.10\leq x/c\leq0.40$ within two observation pixels (\cref{tab:continued-neural-rays}); the cylinder bow shock reproduces the shape of Billig's correlation to within $0.023D$ with a stand-off 8 to 12 percent larger, a difference shared by the PE/BIR neural, physical and hybrid traces. CFD and extraction conventions are possible contributors, but correlated extraction errors prevent a unique attribution, and these are not LocalFront measurements.

The five-variant comparisons match data and update count, the U-Net comparison additionally matches parameter count, and shock adaptation, which uses expanded supervision and additional updates, is evaluated on its own.

The expanding-flow branch is trained on analytical fields and evaluated on the CFD cases as a region detector, with the centred-fan geometry supplied by the analytical construction of \cref{sec:final-multistructure}. Contact surfaces, sonic lines, jets and separation or reattachment boundaries are natural candidates for further learned classes within the same task-preserving framework.

The analytic vortex benchmark provides a label-independent test of the retained task, with a prescribed core definition and no shock interaction. Before adaptation, HJ and U-Net reach $0.672\pm0.086$ and $0.600\pm0.080$ against the prescribed peak-speed-radius disk; U-Net recovers every centre, one HJ seed misses all 16 $M=0.3$ states, and a fixed swirling-strength threshold recovers the clean disks almost exactly (Supplementary Section S3). The benchmark therefore measures retention, not vortex accuracy in the presence of shocks, and a controlled shock--vortex interaction with a numerically verified reference is the natural next test. Repeated states are correlated observations, not additional independent geometries, and a prospective claim applies only to a model that was fixed before the case was inspected.

The equal-budget corrected-target experiment directly tests whether shock adaptation preserves the other task. Joint HJ reaches the largest corrected-target shock Dice and the shared-decoder U-Net nearly matches restricted HJ on the airfoil, but both unprotected adaptations alter the retained core output, often substantially. A retention penalty recovers most but not all of it; restricted HJ alone gives exact-zero probability drift on all 129 seed--field evaluations. Total capacity is matched between HJ and U-Net, whereas the number and topology of trainable parameters differ, so the experiment tests task isolation under a common data and update budget rather than every architectural degree of freedom. The comparison distinguishes exact output preservation from corrected-target agreement and independent geometric accuracy.

\section{Conclusions}
\label{sec:conclusions}
We have developed a physics-audited multi-label formulation for the joint segmentation of shocks and vortex cores in compressible CFD fields. Independent decoders represent overlapping structures, branch-restricted adaptation gives the protected task an exact preservation property, and learned maps, physical proposals and hybrid outputs stay separate throughout the evaluation.

The airfoil and cylinder results show several practical consequences of this formulation. Finite-width shock envelopes are substantially more stable across the paired airfoil grids than one-pixel ridges, with median Dice 0.8104 against 0.5569, so they carry the stability measures while the ridge is kept for point-distance evaluation. Physical rotational inputs improve agreement with the vortex reference on the airfoil, whereas neither gradient projection nor hybrid support gives a uniform advantage across tasks and geometries, and the capacity-matched comparison before adaptation is unresolved within seed variability. Expanded shock supervision combined with branch-restricted adaptation increases front coverage while leaving every vortex output unchanged on 86 evaluation fields and four controls, and native-precision replay reproduces the complete learned output on all 86 fields. Against references independent of the training labels, the frozen shock-adapted models cover the five SU2 ray segments over $0.10\leq x/c\leq0.40$ within two pixels, with a mean normal displacement of about 0.003 chord, and the PE/BIR cylinder traces reproduce the shape of the Billig bow-shock correlation to within 0.019 to 0.023 diameters, with a stand-off excess of 8 to 12 percent that is shared by all extraction methods and decreases under grid refinement.

The four-arm equal-update control is the central result.

With the same corrected targets and 900 additional updates, restricted HJ and the shared-decoder U-Net reach airfoil shock Dice $0.936\pm0.004$ and $0.935\pm0.010$ and joint HJ $0.958\pm0.001$, but the corresponding airfoil core Dice are $0.827\pm0.048$, $0.083\pm0.074$ and $0.382\pm0.382$. Only restricted HJ is bitwise invariant on all 129 seed--field evaluations, and on the independent analytic vortex fields its disk Dice stays at $0.672\pm0.086$ while the U-Net falls from $0.600\pm0.080$ to $0.081\pm0.066$. A U-Net with a fixed parent-logit retention penalty recovers airfoil core Dice $0.775\pm0.116$ and analytic-disk Dice $0.579\pm0.047$. Thus, the penalty recovers most of the lost agreement, while structural restriction preserves the full output exactly. These results support task isolation when one structure must be refined while another is preserved; they do not imply a uniform architectural advantage across shock and vortex measures.

The extended representation assigns distinct outputs to vortex-core candidates, wake/shear regions and expanding-flow regions, and the comparison with ideal shock--expansion theory separates expanding-flow regions from centred Prandtl--Meyer fans, which are supplied by the analytical construction. Three extensions follow directly from these results: a controlled shock--vortex interaction with a numerically verified reference, which would turn the retained-task test into an accuracy test after interaction; a full-front precision and localization evaluation on the expanded shock reference; and learned classes for contact surfaces, sonic lines and separation boundaries within the same task-preserving framework.

\input{data_availability}

\section*{CRediT authorship contribution statement}
Ehsan Roohi: Conceptualization, Methodology, Software, Investigation, Data curation, Validation, Visualization, Writing (original draft), Writing (review and editing).

\section*{Declaration of competing interest}
The author declares no known competing financial interests or personal relationships that could have appeared to influence the work reported in this paper.

\section*{Acknowledgments}
The author gratefully acknowledges the Unity high-performance computing cluster at the University of Massachusetts Amherst for providing the computational resources used in this study.
\section*{Supplementary material}
The accompanying Supplementary PDF contains the full CFD formulation and numerical settings, the complete neural architecture and optimization specification, all validation tables, extended transfer and ablation figures, replay checks, provenance records and limitations. Full-resolution time-resolved movies are distributed separately from the Overleaf source and are linked through the public research repository.

\section*{Declaration of generative AI and AI-assisted technologies}
During manuscript preparation the author used generative-AI tools for language editing, code assistance and consistency checks. The author reviewed the resulting text, code and figures and remains responsible for the scientific content.

\bibliographystyle{elsarticle-num}
\bibliography{references}
\end{document}

%% file: landscapefig.tex
\RequirePackage{atbegshi}
\RequirePackage{refcount}
\RequirePackage{calc}
\ifdefined\pdfpagewidth\else
  \ifdefined\pagewidth\let\pdfpagewidth\pagewidth\let\pdfpageheight\pageheight\fi
\fi
\newcounter{lsfigcount}
\newif\iflsthispage
\newlength{\lsmargin}
\newlength{\lstextwidth}
\newlength{\lstextheight}
\newlength{\lsfigureheight}
\newlength{\lsshift}                                      
\makeatletter
\newcommand{\ls@mark}{\stepcounter{lsfigcount}%
  \protected@write\@auxout{}{\string\newlabel{lsfig@\thelsfigcount}{{}{\thepage}}}}
\newenvironment{landscapefigure}[1][]{%
  \begin{figure}[p]#1%
  \ls@mark%
  \setlength{\lsshift}{\dimexpr 1in+\hoffset+\oddsidemargin-\lsmargin\relax}%
  \noindent\hbox to\textwidth\bgroup\hspace*{-\lsshift}%
  \begin{minipage}[t][\textheight][t]{\lstextwidth}\centering}%
 {\end{minipage}\hss\egroup\end{figure}}
\newcommand{\ls@check}{%
  \global\lsthispagefalse
  \@tempcnta=\@ne
  \loop\ifnum\@tempcnta<80
    \ifcsname r@lsfig@\the\@tempcnta\endcsname
      \edef\ls@pg{\getpagerefnumber{lsfig@\the\@tempcnta}}%
      \ifnum\ls@pg=\value{page}\relax\global\lsthispagetrue\fi
    \fi
    \advance\@tempcnta\@ne
  \repeat}
\AtBeginShipout{%
  \ls@check
  \iflsthispage
    \pdfpagewidth=\paperheight \pdfpageheight=\paperwidth
    \setlength{\@tempdima}{\dimexpr 1in+\voffset+\topmargin+\headheight+\headsep-\lsmargin\relax}%
    \setbox\AtBeginShipoutBox=\vbox{\kern-\@tempdima\box\AtBeginShipoutBox}%
    \AtBeginShipoutUpperLeftForeground{%
      \raisebox{-\dimexpr\paperwidth-8mm\relax}[0pt][0pt]{\makebox[\paperheight][c]{\normalfont\normalsize\thepage}}}%
  \else
    \pdfpagewidth=\paperwidth \pdfpageheight=\paperheight
  \fi}
\makeatother

%% file: v4_results.tex
\input{v4_contrast_macros}
\subsection{Effect of architecture, differential inputs and gradient projection}
\label{sec:v4-results}
\Cref{tab:v4-test-airfoil,tab:v4-test-cylinder} summarize the five-model comparison over three independent initialization seeds. All variants see the same frames within each geometry family and exactly the same 2,400 training patches and 600 optimizer updates. The reported mean therefore describes the average over three training initializations, while the sample standard deviation (SD) shows how strongly the result changes with initialization. Validation-set results and the individual seed values are given in the Supplementary material.
\input{v4_tables_test}

The three-seed experiment does not resolve a consistent benefit from gradient projection. The joint-minus-no-PCGrad ML-only shock-Dice difference is \VFourPcgradAirfoil{} for the airfoil and \VFourPcgradCylinder{} for the cylinder. Thus conflicting shock and vortex gradients are present in the shared encoder, but PCGrad does not improve every reported measure. By contrast, removing the \emph{differential inputs} sharply reduces airfoil core agreement. These inputs are local velocity-gradient and rotation-derived channels rather than additional primitive variables; they help distinguish a compact rotating core from a broad shear layer. Because some of the same diagnostics also contribute to the weak vortex reference, the magnitude of this gain is interpreted together with the full-field probability maps rather than as independent ground-truth accuracy.

The transformer-based model produces higher airfoil component F1 than the joint model at the selected operating points but lower core pixel Dice, so component recovery and pixel coverage rank these outputs differently. SegFormer-B0 contains 3,721,958 parameters against 268,496 for the joint model. The fixed-update experiment compares training configurations under a common budget; the capacity-matched comparison below additionally controls parameter count, though not every architectural choice.

\subsection{Capacity-matched conventional U-Net comparison}
\label{sec:capacity-unet}
We compare a conventional shared-encoder, shared-decoder U-Net with the branch-specialized harmonized model using the same seven channels, patch bank, three initialization seeds, batch size four, six-head objective, optimizer schedule and 600 updates per seed. Its 267,639 parameters differ by only 0.319\% from the 268,496 parameters of the primary model. It omits both the branch-specific decoders and PCGrad, so the comparison isolates the architecture package as a whole; the separate no-PCGrad experiment above addresses gradient projection on its own. All six foreground and auxiliary logits remain independent in both models. \Cref{tab:capacity-unet} gives the scores.

\begin{table}[!htbp]
\centering\small
\caption{Capacity-matched comparison before shock adaptation. The branch-specialized model and conventional U-Net have nearly the same total parameter count and use identical inputs, data and update budgets. Shock and core Dice are pixel-overlap scores against the original physics-derived weak references, averaged first within each geometry family and then across the three seeds; $\pm$ denotes sample SD. The airfoil shock reference is intentionally sparse, so its Dice score measures agreement with that sparse target rather than recovery of the entire visible front.}
\label{tab:capacity-unet}
\begin{tabularx}{\textwidth}{Y r r r}
\toprule
Model & Parameters & Shock Dice & Core Dice\\
\midrule
Branch-specialized & 268,496 & $0.890\pm0.100$ & $0.527\pm0.015$\\
Conventional U-Net & 267,639 & $0.806\pm0.129$ & $0.533\pm0.019$\\
\bottomrule
\end{tabularx}
\end{table}
The paired shock-Dice differences are $+0.252$, $-0.036$ and $+0.034$ across seeds (two-sided paired $t$ test, $p=0.437$, $n=3$). With three seeds the comparison resolves neither a difference nor an equivalence between the two architectures, and we draw no architectural ranking from it. Core pixel Dice agrees to within 0.006. These scores measure agreement with the stored targets. The same-field maps in \cref{fig:capacity-fields} show what they do and do not capture: before shock adaptation neither shock head follows the long airfoil fronts, whereas both find the cylinder bow shock, and the branch-specialized core probability also includes a continuous wake band rather than only compact cores. Because the later shock adaptation changes both supervision and budget, its equal-additional-budget HJ/U-Net/joint-HJ control is reported separately in \cref{sec:corrected-target-control}.

Object F1 is omitted from this comparison because the ranking depends on the component convention: the earlier evaluation convention gives 0.6004 versus 0.6187 (HJ/U-Net), whereas component identities restricted to visible support give 0.4534 versus 0.4390. Both values are reported in the Supplementary material, and neither ordering is used to claim superiority.

\begin{landscapefigure}
\includegraphics[width=\linewidth,height=\lsfigureheight,keepaspectratio]{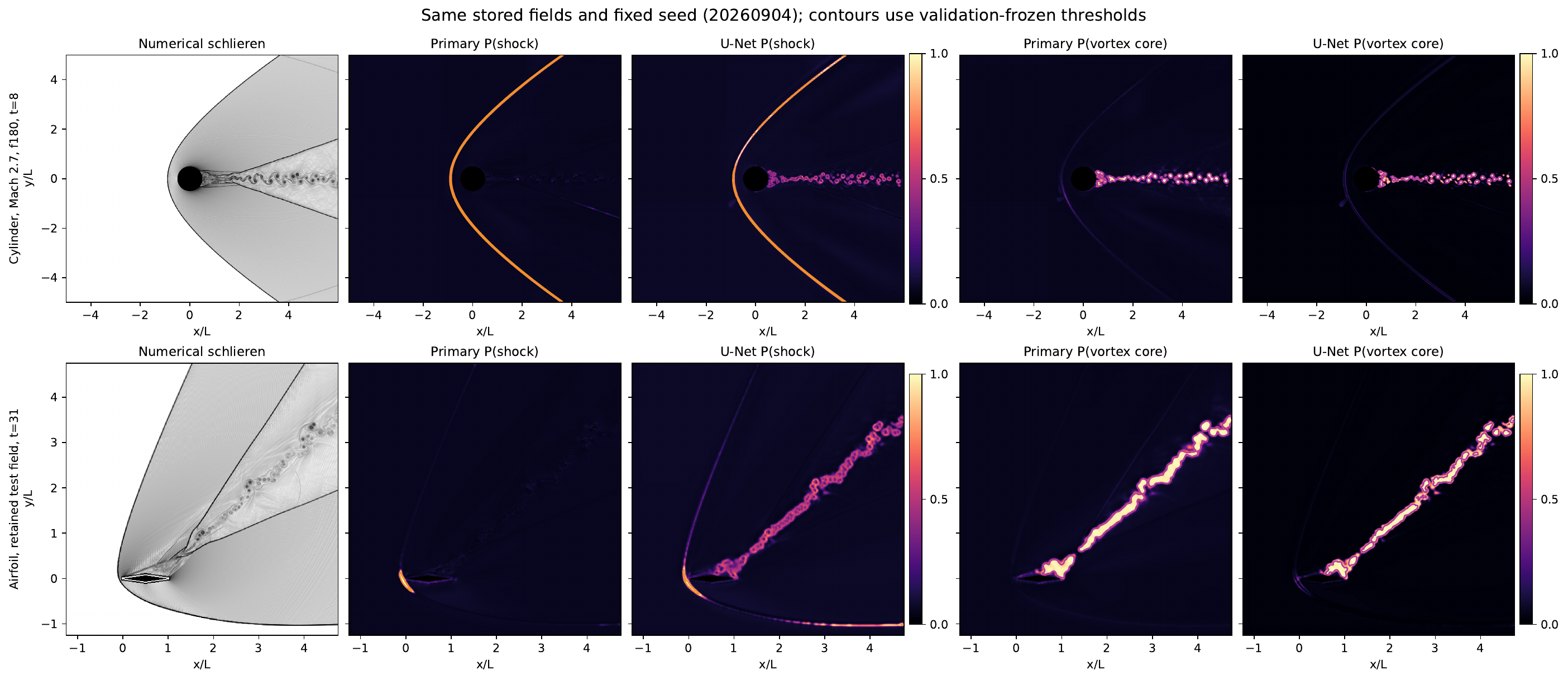}
\caption{Same-field probability maps for the capacity-matched comparison before shock adaptation. Each airfoil and cylinder field is shown to both networks with identical seven-channel inputs; only the architecture differs. Shock probability maps show that neither pre-adaptation model follows the full long airfoil fronts, although both detect the cylinder bow shock. The vortex/core map of the branch-specialized model also contains an extended wake band, illustrating that a high weak-reference score does not imply that every positive pixel is a compact vortex core. The panels are therefore used to interpret the tabulated Dice scores spatially, not as a visual ranking by contrast.}
\label{fig:capacity-fields}
\end{landscapefigure}
\FloatBarrier

\subsection{Contribution of physical support}
For the joint model, hybrid-minus-ML shock Dice is \VFourHybridShockAirfoil{} on the airfoil and \VFourHybridShockCylinder{} on the cylinder, and the corresponding changes in core-component F1 are \VFourHybridObjectAirfoil{} and \VFourHybridObjectCylinder{}. Physical support therefore acts as a selective addition to the learned mask, not as a uniformly beneficial correction. It can extend a broken front, but the added pixels can also reduce precision or connect two nearby core candidates that would otherwise count as separate objects. The shock hybrid uses a dilated version of the same physical proposal that contributes to the weak reference; for that reason its Dice is interpreted as \emph{reference-conditioned agreement}, not as an independent accuracy estimate. \Cref{fig:v4-airfoil-controls,fig:v4-cylinder-controls} place all learned variants, the hybrid additions and the weak reference on the same fields so that these effects are visible.

\input{v4_figure_captions}
\begin{landscapefigure}
\includegraphics[width=\linewidth,height=\lsfigureheight,keepaspectratio]{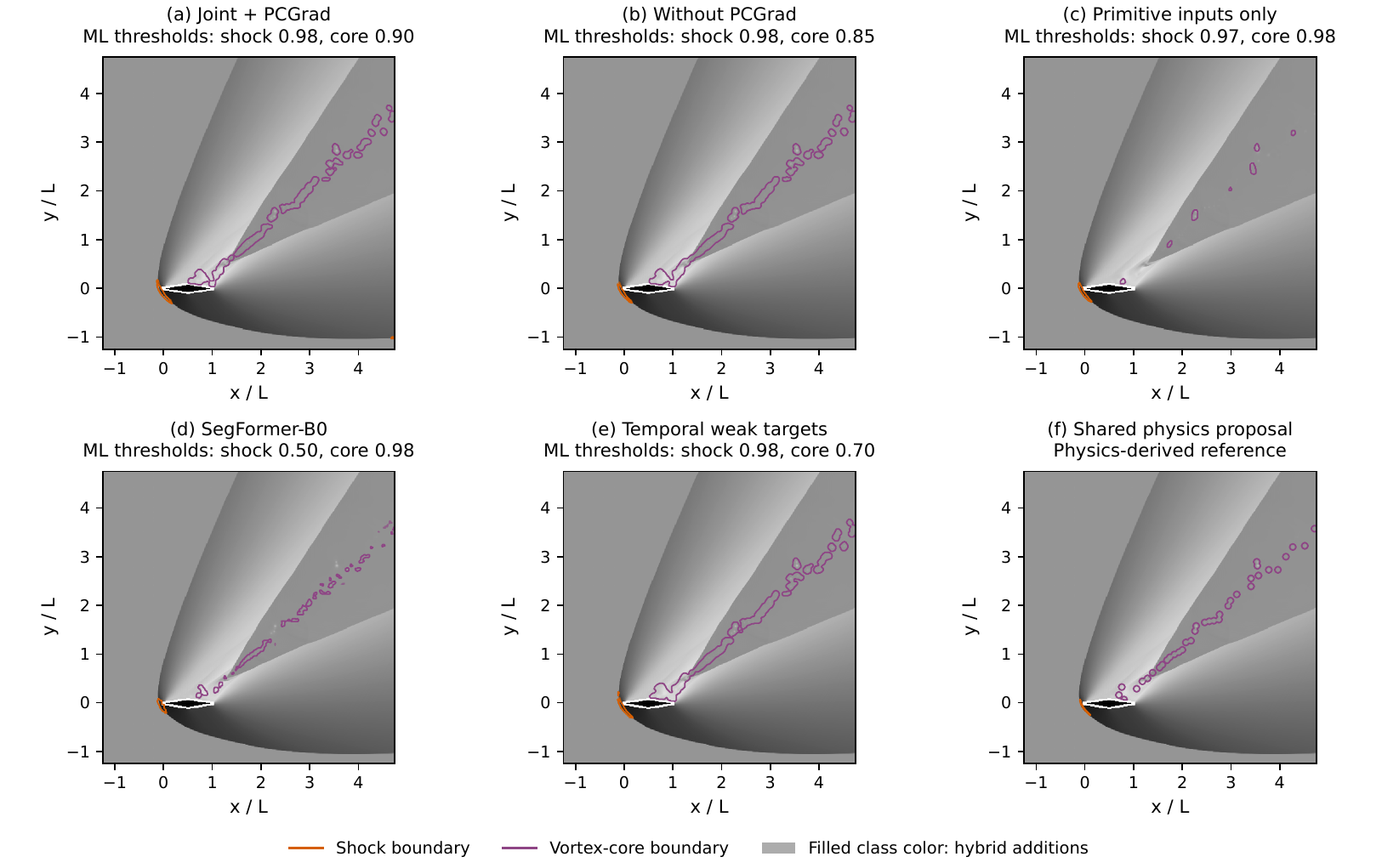}
\caption{\VFourAirfoilComparisonCaption}
\label{fig:v4-airfoil-controls}
\end{landscapefigure}

\begin{landscapefigure}
\includegraphics[width=\linewidth,height=\lsfigureheight,keepaspectratio]{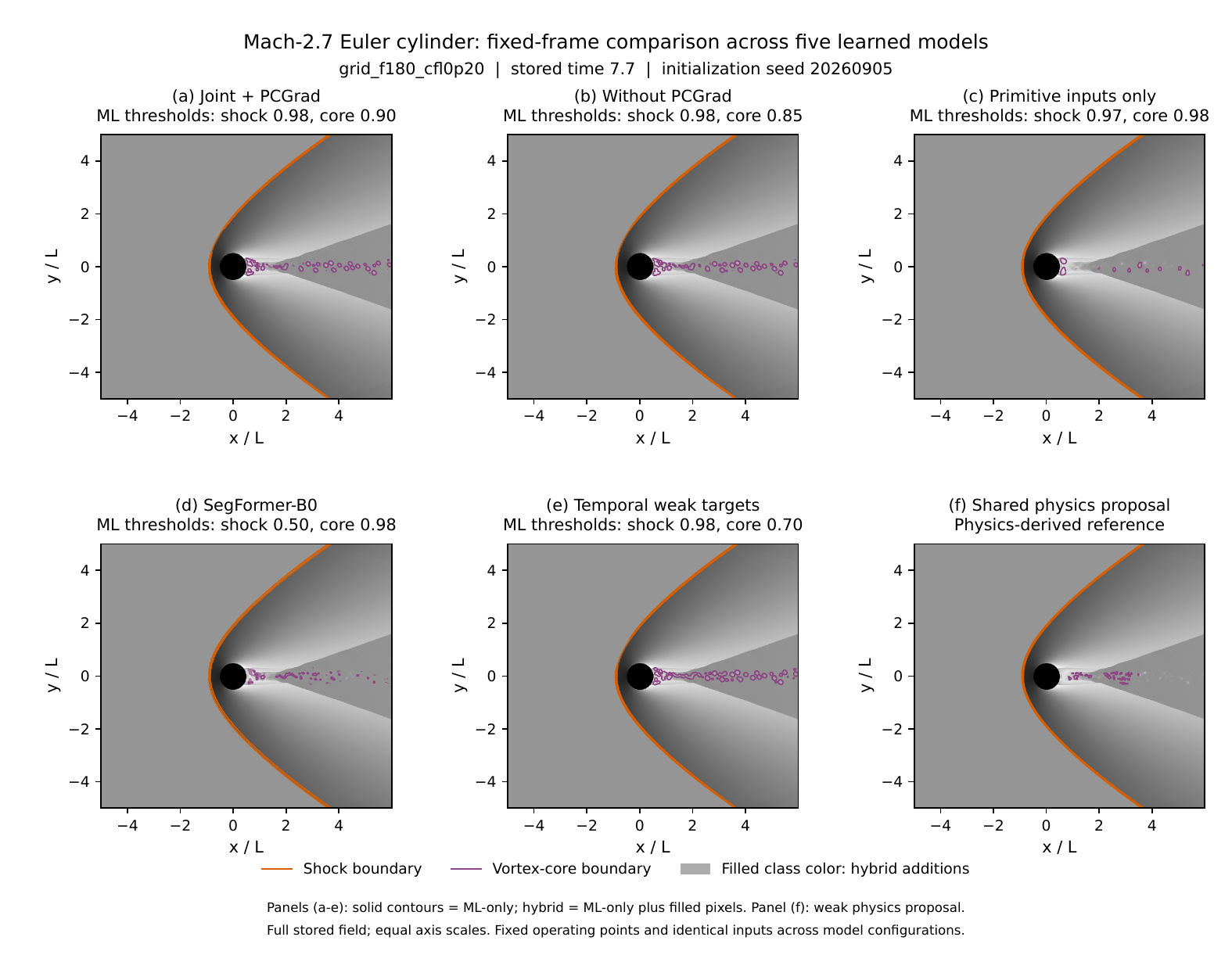}
\caption{\VFourCylinderComparisonCaption}
\label{fig:v4-cylinder-controls}
\end{landscapefigure}
\FloatBarrier

\subsection{Front completeness and task-preserving adaptation}
\label{sec:v4-repair}
The Reynolds-$10^6$ airfoil field makes a key distinction visible: a model can agree well with a sparse reference while still missing most of the physical front. The original weak shock reference contains only 48 positive pixels in this field, whereas the expanded full-front teacher contains 5,742. Of those expanded positives, 5,582 lie in locations that the sparse reference had labelled as background once unobserved padding is excluded. The same conflict appears across the 84 airfoil training frames: the sparse reference contains 5,006 positive pixels, the expanded teacher 493,643, and 478,403 of the expanded positives had previously received negative supervision. The missing front is therefore primarily a target-definition problem, not a visualization-threshold problem. Shock adaptation changes the supervision so that the model is allowed to learn the complete front; simply lowering a display threshold would not remove the contradictory labels.

\Cref{tab:v4-repair} compares the base and noise-augmented shock-adapted models against the same expanded reference and ignore policy. These values use the expanded reference and are therefore reported separately from the sparse-reference architecture comparison. The full-field overlays in \cref{fig:v4-repair-airfoil,fig:v4-repair-cylinder} show the recovery of the extended airfoil fronts and the preservation of the cylinder bow front; compact shock responses also appear within parts of the wake, where compression and rotational structures coexist.
\input{v4_table_repair}

Task preservation holds exactly on all 86 evaluation fields and four controls: vortex probabilities and both ML-only and hybrid vortex masks are unchanged, while 20 shock state tensors change. On the thermodynamically flat subset of the velocity-noise control, the clean-target adaptation produces 11,267 shock-positive pixels, whereas the noise-augmented adaptation produces none. The unadapted model also produces none there, but it misses much of the extended front. Suppression of free-stream responses and recovery of shock extent therefore have to be assessed together, and the noise-augmented configuration achieves both. The frozen vortex pathway preserves its original output throughout these changes.

\begin{landscapefigure}
\includegraphics[width=\linewidth,height=\lsfigureheight,keepaspectratio]{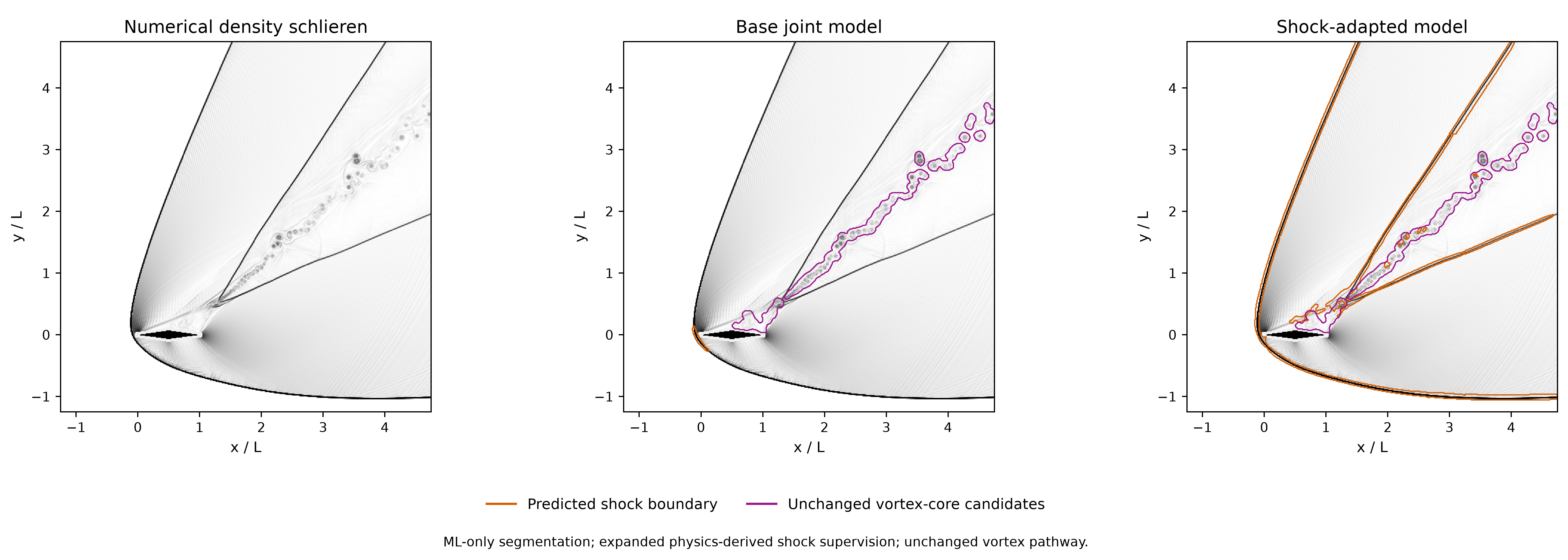}
\caption{Effect of task-preserving shock adaptation on the full $M_\infty=3$, $\alpha=40\dg$, $\Rey_c=10^6$ airfoil field at $t=30.7$. The left panel is numerical density schlieren; the middle panel is the base HJ prediction; the right panel is the shock-adapted prediction. Orange contours are learned shock boundaries and purple contours are vortex-core boundaries. The important comparison is that the adapted model extends the orange shock response along the previously missed fronts while the purple vortex response is exactly unchanged. Schlieren uses the fixed mapping $\exp[-0.08(L/\rho_\infty)|\nabla\rho|]$, with $L=c$ for an airfoil and $L=D$ for a cylinder, so all panels use the same density-gradient contrast. Equal physical axis scales and the complete stored field are retained.}
\label{fig:v4-repair-airfoil}
\end{landscapefigure}
\begin{landscapefigure}
\includegraphics[width=\linewidth,height=\lsfigureheight,keepaspectratio]{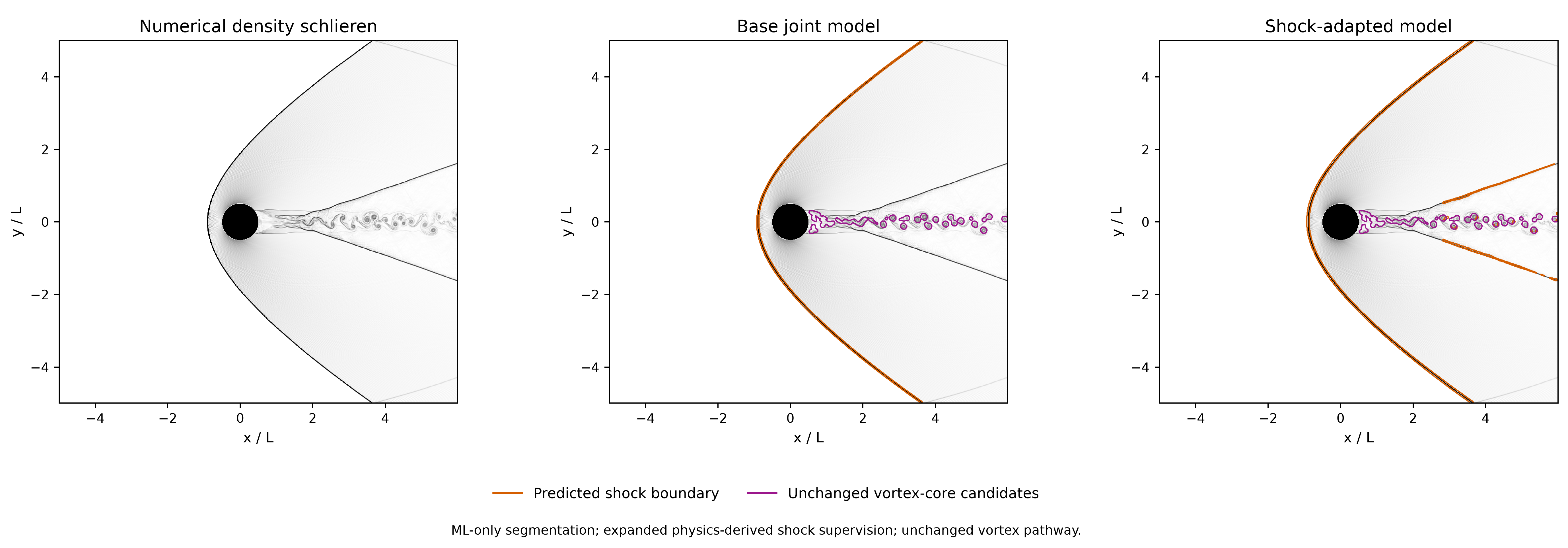}
\caption{The same base-versus-adapted comparison on the $M_\infty=2.7$ circular-cylinder run with 180 cells per diameter (f180), CFL 0.20 and $t=7.7$. Orange denotes the learned shock boundary and purple the vortex-core response; numerical schlieren uses the same fixed transfer function as \cref{fig:v4-repair-airfoil}. The detached bow shock is already present before adaptation and remains in the same location. Adaptation mainly extends downstream compression-front coverage beginning about $1.5D$ behind the cylinder, while the complete purple vortex probability field remains unchanged. Equal axis scales preserve the circular body and make geometric displacement directly visible.}
\label{fig:v4-repair-cylinder}
\end{landscapefigure}
\FloatBarrier

%% file: v4_contrast_macros.tex
\newcommand{\VFourPcgradAirfoil}{$-0.0395\pm0.1481$}
\newcommand{\VFourHybridShockAirfoil}{$-0.1538\pm0.0735$}
\newcommand{\VFourHybridObjectAirfoil}{$-0.0063\pm0.0064$}
\newcommand{\VFourPcgradCylinder}{$+0.0014\pm0.0084$}
\newcommand{\VFourHybridShockCylinder}{$+0.0851\pm0.0924$}
\newcommand{\VFourHybridObjectCylinder}{$-0.0058\pm0.0160$}

%% file: v4_tables_test.tex
\begin{table}[!htbp]
\centering\small
\caption{Airfoil development-test agreement with the original sparse physics-derived weak reference: 20 frames per seed and 943 positive shock-reference pixels in total. ``Sparse-ref. Dice'' is pixel overlap with that shock reference, ``Core Dice'' is pixel overlap with the weak vortex-core reference, and ``Core object F1'' measures one-to-one recovery of connected core objects. These scores do not measure complete shock-front recovery because the original shock target marks only a small part of the visible front; completeness is evaluated later with the expanded target. Values are mean $\pm$ sample SD over three initializations. ``ML'' means learned probabilities only, ``Hybrid'' includes the fixed physical-support additions, and the $\tau=0.9$ rows apply the same probability threshold to both learned classes. Physics-only self-agreement is omitted.}
\label{tab:v4-test-airfoil}
\setlength{\tabcolsep}{3pt}
\begin{tabularx}{\textwidth}{Y l r r r}
\toprule
Variant & Branch & Sparse-ref. Dice & Core Dice & Core object F1\\
\midrule
Joint & ML & $0.8717\pm0.1094$ & $0.8269\pm0.0477$ & $0.5896\pm0.1910$\\
Joint & Hybrid & $0.7178\pm0.1828$ & $0.8244\pm0.0498$ & $0.5833\pm0.1851$\\
Joint & ML, $\tau=0.9$ & $0.6611\pm0.3534$ & $0.8679\pm0.0059$ & $0.5467\pm0.0985$\\
\addlinespace[3pt]
No PCGrad & ML & $0.9112\pm0.0387$ & $0.8231\pm0.0516$ & $0.5874\pm0.1781$\\
No PCGrad & Hybrid & $0.7134\pm0.1201$ & $0.8207\pm0.0531$ & $0.5808\pm0.1724$\\
No PCGrad & ML, $\tau=0.9$ & $0.6093\pm0.3333$ & $0.8695\pm0.0096$ & $0.5543\pm0.0834$\\
\addlinespace[3pt]
Primitives & ML & $0.8554\pm0.0853$ & $0.1379\pm0.0761$ & $0.2384\pm0.1196$\\
Primitives & Hybrid & $0.7426\pm0.1019$ & $0.1378\pm0.0759$ & $0.2380\pm0.1189$\\
Primitives & ML, $\tau=0.9$ & $0.7781\pm0.1047$ & $0.3330\pm0.3049$ & $0.3277\pm0.2643$\\
\addlinespace[3pt]
SegFormer-B0 & ML & $0.8277\pm0.0507$ & $0.5131\pm0.0794$ & $0.6515\pm0.0473$\\
SegFormer-B0 & Hybrid & $0.8401\pm0.0529$ & $0.5117\pm0.0753$ & $0.6285\pm0.0499$\\
SegFormer-B0 & ML, $\tau=0.9$ & $0.7973\pm0.0137$ & $0.7285\pm0.0179$ & $0.6754\pm0.0582$\\
\addlinespace[3pt]
Temporal targets & ML & $0.9075\pm0.0731$ & $0.8404\pm0.0335$ & $0.5974\pm0.1273$\\
Temporal targets & Hybrid & $0.7224\pm0.1896$ & $0.8380\pm0.0353$ & $0.5925\pm0.1239$\\
Temporal targets & ML, $\tau=0.9$ & $0.6880\pm0.2897$ & $0.8705\pm0.0127$ & $0.5913\pm0.0653$\\
\bottomrule
\end{tabularx}
\end{table}
\FloatBarrier

\begin{table}[!htbp]
\centering\small
\caption{Cylinder development-test agreement with the original physics-derived weak reference: 23 frames per seed. The three reported quantities have the same meanings as in the airfoil table: shock pixel Dice, vortex-core pixel Dice and object-level core F1. Values are mean $\pm$ sample SD over three initializations. ``ML'' is the learned mask alone, ``Hybrid'' adds physically supported pixels, and the $\tau=0.9$ rows use a common learned probability threshold of 0.9 for both classes. Physics-only self-agreement is omitted.}
\label{tab:v4-test-cylinder}
\setlength{\tabcolsep}{3pt}
\begin{tabularx}{\textwidth}{Y l r r r}
\toprule
Variant & Branch & Sparse-ref. Dice & Core Dice & Core object F1\\
\midrule
Joint & ML & $0.9078\pm0.0926$ & $0.2278\pm0.0204$ & $0.3172\pm0.0452$\\
Joint & Hybrid & $0.9929\pm0.0007$ & $0.2572\pm0.0324$ & $0.3114\pm0.0382$\\
Joint & ML, $\tau=0.9$ & $0.9827\pm0.0058$ & $0.2237\pm0.0310$ & $0.3167\pm0.0364$\\
\addlinespace[3pt]
No PCGrad & ML & $0.9063\pm0.0842$ & $0.2256\pm0.0048$ & $0.2959\pm0.0685$\\
No PCGrad & Hybrid & $0.9927\pm0.0012$ & $0.2412\pm0.0168$ & $0.2948\pm0.0625$\\
No PCGrad & ML, $\tau=0.9$ & $0.9818\pm0.0072$ & $0.2259\pm0.0213$ & $0.3088\pm0.0332$\\
\addlinespace[3pt]
Primitives & ML & $0.9452\pm0.0352$ & $0.1664\pm0.0694$ & $0.1864\pm0.0330$\\
Primitives & Hybrid & $0.9905\pm0.0034$ & $0.1639\pm0.0660$ & $0.1855\pm0.0335$\\
Primitives & ML, $\tau=0.9$ & $0.9775\pm0.0051$ & $0.1391\pm0.0258$ & $0.1985\pm0.0344$\\
\addlinespace[3pt]
SegFormer-B0 & ML & $0.9586\pm0.0494$ & $0.2477\pm0.0155$ & $0.3241\pm0.0187$\\
SegFormer-B0 & Hybrid & $0.9691\pm0.0312$ & $0.2965\pm0.0020$ & $0.3365\pm0.0248$\\
SegFormer-B0 & ML, $\tau=0.9$ & $0.9209\pm0.0420$ & $0.2680\pm0.0150$ & $0.2900\pm0.0435$\\
\addlinespace[3pt]
Temporal targets & ML & $0.9033\pm0.0745$ & $0.2476\pm0.0407$ & $0.2702\pm0.0311$\\
Temporal targets & Hybrid & $0.9894\pm0.0028$ & $0.2478\pm0.0417$ & $0.2784\pm0.0371$\\
Temporal targets & ML, $\tau=0.9$ & $0.9776\pm0.0111$ & $0.2244\pm0.0147$ & $0.2639\pm0.0661$\\
\bottomrule
\end{tabularx}
\end{table}
\FloatBarrier

%% file: v4_figure_captions.tex
\newcommand{\VFourAirfoilComparisonCaption}{Full-field comparison of the five learned variants on the same $M_\infty=3$, $\alpha=40\dg$, $\Rey_c=10^6$, $t=30.7$ airfoil state (second initialization seed), before the later shock-adaptation stage. Panels (a)--(e) are the learned variants and panel (f) is the physics-derived weak reference. Orange contours are machine-learning-only (ML-only) shock boundaries and purple contours are ML-only vortex-core boundaries; filled regions in the same colors are pixels added by the hybrid physical-support rule. The shock/core numbers printed above each panel are the fixed validation-selected probability thresholds used to convert those probability maps to contours. The grayscale background is normalized log pressure, $\operatorname{clip}(\log(p/p_\infty),-6,6)/3$, displayed with grayscale limits $[-1,1]$; transformed values outside that display interval are saturated rather than rescaled panel by panel. At this stage both the learned models and the sparse weak reference describe mainly the leading-edge portions of the airfoil shocks, so this figure compares model behavior under the original target definition rather than full-front accuracy. The adapted full-front result is shown in \cref{fig:v4-repair-airfoil}. All panels use the same flow field and equal coordinate scales.}
\newcommand{\VFourCylinderComparisonCaption}{Full-field comparison of the same five learned variants on the $M_\infty=2.7$ Euler/slip circular-cylinder case with 180 cells per diameter (f180), CFL 0.20 and $t=7.7$ (second initialization seed). Panel order and color coding follow \cref{fig:v4-airfoil-controls}: orange is shock, purple is vortex core, contours are ML-only predictions and filled additions are supplied by the hybrid physical-support rule. The numbers above the panels are the same type of fixed validation-selected shock/core probability thresholds as in the airfoil figure. The normalized log-pressure background and its fixed display limits are identical to those used for the airfoil comparison. Equal coordinate scales preserve the circular body. The panels show that compact rotational candidates and the broader connected wake response are distinct spatial behaviors even when both contribute to a vortex-related probability map.}

%% file: v4_table_repair.tex
\begin{table}[!htbp]
\centering\small
\caption{Shock-adaptation comparison on 43 development-test frames (20 airfoil and 23 cylinder). Every row is scored against the same expanded physics-derived shock reference, which covers the intended full front and therefore differs from the sparse reference used in the earlier architecture comparison. Dice measures total mask overlap, precision is the fraction of predicted shock pixels inside the expanded reference, and recall is the fraction of reference shock pixels recovered. ``Base ML'' is the pre-adaptation learned mask, ``Adapted ML'' is the task-preserving learned mask after shock refinement, and ``Adapted hybrid'' adds the fixed physical support. The vortex pathway is frozen throughout.}
\label{tab:v4-repair}
\begin{tabular}{llrrr}
\toprule
Family & Shock branch & Dice & Precision & Recall\\
\midrule
Airfoil & Base ML & 0.0428 & 1.0000 & 0.0218\\
Airfoil & Adapted ML & 0.9345 & 0.9560 & 0.9139\\
Airfoil & Adapted hybrid & 0.9350 & 0.9560 & 0.9149\\
Cylinder & Base ML & 0.9587 & 1.0000 & 0.9206\\
Cylinder & Adapted ML & 0.9633 & 0.9346 & 0.9938\\
Cylinder & Adapted hybrid & 0.9642 & 0.9327 & 0.9979\\
\bottomrule
\end{tabular}
\end{table}
\FloatBarrier

%% file: corrected_control_extension.tex
\subsection{Equal-budget corrected-target control and analytic vortex reference}
\label{sec:corrected-target-control}
The corrected shock supervision is tested in a three-seed control designed to remove the unequal adaptation budget identified in \cref{sec:capacity-unet}. Here an \emph{arm} means one adaptation strategy applied to the same parent model under the same data schedule. Every arm starts from its own 600-update parent and then receives exactly 600 clean shock-target updates followed by 300 velocity-noise-augmented updates. Batch size, $128\times128$ training patches, AdamW learning-rate schedules and validation-only threshold selection are identical. The comparison therefore asks what changes when the \emph{allowed trainable pathway} changes, rather than when one model simply receives more optimization.

Four adaptation strategies are compared. In task-preserving HJ, only 75,891 shock-decoder parameters are trainable; the shared encoder and the complete vortex pathway are frozen. In the 267,639-parameter conventional U-Net, 107,982 parameters in the shared decoder and shock projection are trainable. Rows of the final projection associated with nonshock outputs are restored after every update, but the shared decoder feeding those rows can still change, so the vortex computation is not structurally isolated. A third, joint-HJ control trains all 268,496 HJ parameters and therefore provides no protected pathway.

The fourth strategy is the same shared-decoder U-Net with a \emph{soft retention penalty}. In addition to shock loss, it penalizes the mean-squared difference between the current vortex logits and the logits produced by the frozen parent on the same training input, with fixed weight $\eta=0.2$. This encourages, but cannot mathematically guarantee, preservation of the parent vortex output. It uses the same parent, trainable parameters and input patches as the unprotected U-Net.

PCGrad is not invoked; gradients of the summed objective are used throughout this control. The extra frozen-parent forward pass means that equal updates do not imply equal floating-point operations (FLOPs) or wall time. The comparison therefore matches inputs, corrected targets, batches and updates, total model capacity is matched only between HJ and U-Net, and trainable capacity is reported rather than claimed equal.

The independent vortex control uses a convected isentropic Euler vortex, a smooth analytical solution that supplies a known vortex location and size without any physics-derived segmentation mask \cite{spiegel2015isentropic}. The vortex centre is $(x_c,y_c)$, $a$ sets its radial length scale, and $\epsilon$ controls its circulation strength. We use the nondimensional radius $s$, defined by $s^2=[(x-x_c)/a]^2+[(y-y_c)/a]^2$, and a calorically perfect gas with ratio of specific heats $\gamma=1.4$. The exact sampled state is
\begin{align}
 f&=\epsilon\exp[(1-s^2)/2], &
 u&=U_\infty-f(y-y_c)/a, & v&=f(x-x_c)/a,\\
 T&=1-\frac{\gamma-1}{2\gamma}\epsilon^2\exp(1-s^2), &
 \rho&=T^{1/(\gamma-1)}, & p&=\rho^\gamma ,
\end{align}
Here $f$ is the local velocity-perturbation amplitude and $T$ is nondimensional temperature. The complete vortex simply translates with the freestream, $x_c(t)=x_c(0)+U_\infty t$. For evaluation we define the analytic core as the disk $s\leq1$, whose radius coincides with the radius of peak tangential speed. This is a deliberately reproducible evaluation definition, not a claim that every CFD vortex has a sharp circular boundary. The grid contains $192\times192$ points on $[-3,3]^2$. We vary freestream Mach number over $M_\infty\in\{0.3,1.5,3\}$, size over $a\in\{0.25,0.5\}$, strength over $\epsilon\in\{0.5,1\}$, and use two initial centres at two times, giving 48 states organized as 24 two-time trajectories. A vortex is counted as centre-recovered if a predicted object contains the known centre or if its centroid lies within $a/2$. Uniform flow, parallel shear, a kinematic potential-strain field and a planar normal shock are negative controls on which no vortex core should be detected. None of these analytic fields is used for training, threshold selection or case-specific tuning.

\Cref{tab:corrected-target-control} collects the scores of the four arms, \cref{fig:corrected-control-airfoil} shows them on a late airfoil field that was chosen before the arms were evaluated, and \cref{fig:matched-control-summary} plots the same quantities with their seed spread.
\input{corrected_target_control_table}

On corrected weak shock references, task-preserving HJ and the shared-decoder U-Net have nearly equal airfoil shock Dice, $0.936\pm0.004$ and $0.935\pm0.010$, while HJ is higher on the cylinder, $0.939\pm0.024$ against $0.895\pm0.022$. Joint HJ obtains the largest shock Dice, $0.958\pm0.001$ and $0.955\pm0.021$, but the retained-task scores show that higher shock agreement can accompany substantial vortex degradation. The task-preserving HJ core probabilities are bitwise identical to their parents on every one of the 129 seed--field evaluations and keep airfoil/cylinder core Dice of $0.827\pm0.048$/$0.228\pm0.020$. The shared-decoder U-Net changes by as much as 0.999 in core probability and falls to $0.083\pm0.074$/$0.001\pm0.001$. Joint HJ also changes the core output, with Dice scores of $0.382\pm0.382$/$0.104\pm0.086$.

The retention-penalized U-Net obtains airfoil shock Dice $0.937\pm0.006$ and keeps airfoil/cylinder core Dice $0.775\pm0.116$/$0.202\pm0.035$; its analytic-disk Dice is $0.579\pm0.047$. The retention penalty thus recovers most of the protected output that unprotected adaptation loses, and we report its residual probability drift alongside the exact invariance of the restricted model. Structural restriction guarantees unchanged outputs for the frozen pathway, whereas a finite penalty trades adaptation against approximation to the parent on the training inputs, and its strength was fixed at $\eta=0.2$ rather than tuned.

Before adaptation, frozen HJ and U-Net have analytic-disk Dice $0.672\pm0.086$ and $0.600\pm0.080$; U-Net recovers all 48 centres in every seed, whereas one HJ seed misses the 16 low-Mach states (\cref{fig:independent-vortex-control}). After shock-only adaptation, the task-preserving HJ result is unchanged, by construction and by direct array comparison. U-Net falls to analytic-disk Dice $0.081\pm0.066$ and mean centre recovery $0.472\pm0.502$, and joint HJ gives $0.593\pm0.192$ and $0.861\pm0.241$. The reported results include all missed cores and the original U-Net false positives, which occupy 6.22\% of the parallel-shear field in one seed. These controls show interference under shared or joint adaptation and exact retained-task isolation under branch restriction. They establish neither a universal HJ advantage nor vortex accuracy after shock interaction.

The three original adapted arms cover the five finite analytical SU2 ray segments. Their conditional mean normal displacements are $0.00284c$ (task-preserving HJ), $0.00303c$ (U-Net) and $0.00247c$ (joint HJ), with mean absolute orientation errors of $0.18\dg$, $0.20\dg$ and $0.15\dg$. These label-independent finite-ray results confirm that the shock gains are not merely a restatement of the corrected masks, but they remain local and do not measure full-front precision.

\begin{landscapefigure}
\includegraphics[width=\linewidth,height=\lsfigureheight,keepaspectratio]{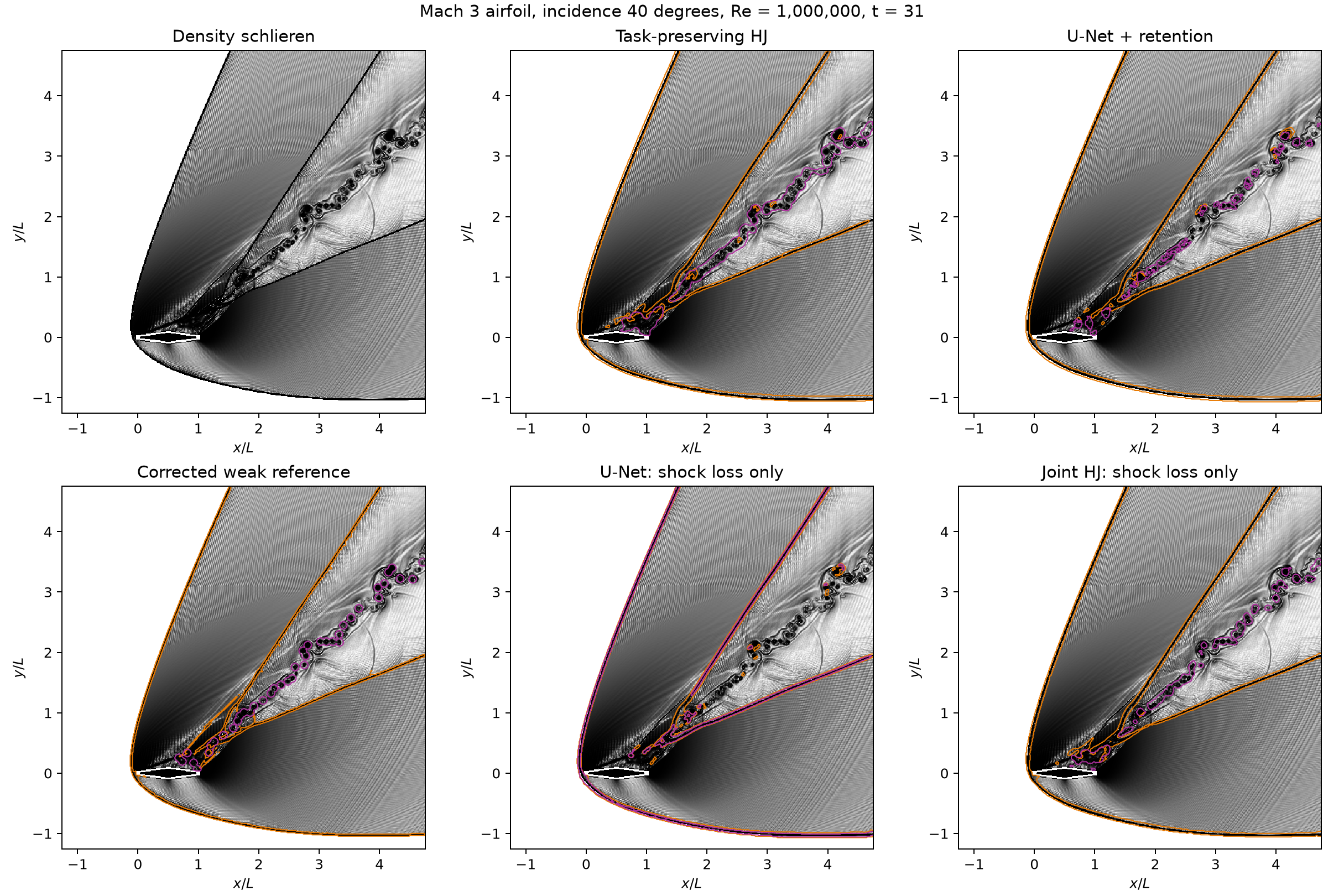}
\caption{Spatial comparison of the four equal-budget adaptation strategies on one late airfoil development field from the first initialization seed; the field was chosen before the four arms were evaluated. Top row: numerical density schlieren, task-preserving HJ, and U-Net with the soft retention penalty, which penalizes mean-squared drift of the vortex logits from the frozen parent with fixed weight $\eta=0.2$. Bottom row: corrected physics-derived weak targets, unprotected shared-decoder U-Net, and fully trainable joint HJ. Orange marks shock masks and purple marks vortex-core masks. The key quantity is not only shock coverage but whether the purple retained task changes while the shock branch is refined: restricted HJ leaves the complete vortex probability array bitwise unchanged, the unprotected U-Net alters it strongly, and the retention penalty reduces but does not eliminate that drift. The reference panel is an algorithmic weak target rather than expert ground truth.}
\label{fig:corrected-control-airfoil}
\end{landscapefigure}

\begin{landscapefigure}
\includegraphics[width=\linewidth,height=\lsfigureheight,keepaspectratio]{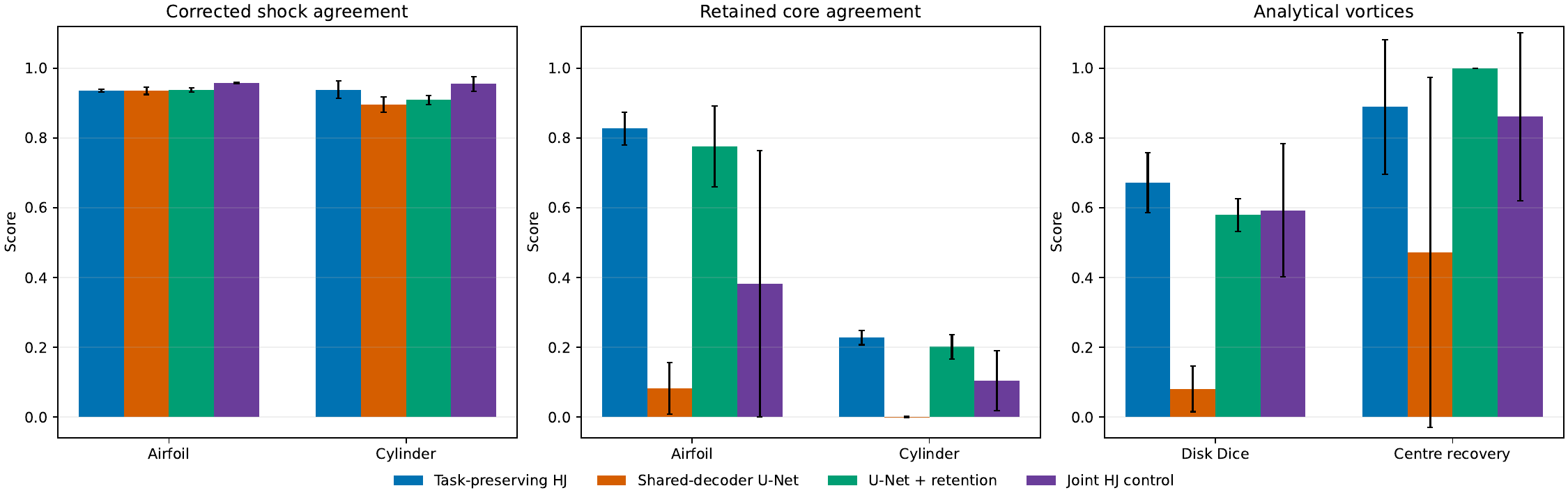}
\caption{Summary of retained-task behavior under the equal 900-update adaptation budget. Each bar is the mean over three initialization seeds and each error bar is the sample SD. Airfoil and cylinder shock/core quantities are Dice scores against corrected physics-derived weak references. Analytic-disk Dice compares the predicted core mask with the prescribed $s\leq1$ disk of the isentropic vortex, while centre recovery is the fraction of analytic vortices whose known centre is captured by a predicted object. The figure therefore juxtaposes target agreement on CFD fields with an independent vortex test whose geometry is known exactly.}
\label{fig:matched-control-summary}
\end{landscapefigure}

\begin{landscapefigure}
\includegraphics[width=\linewidth,height=\lsfigureheight,keepaspectratio]{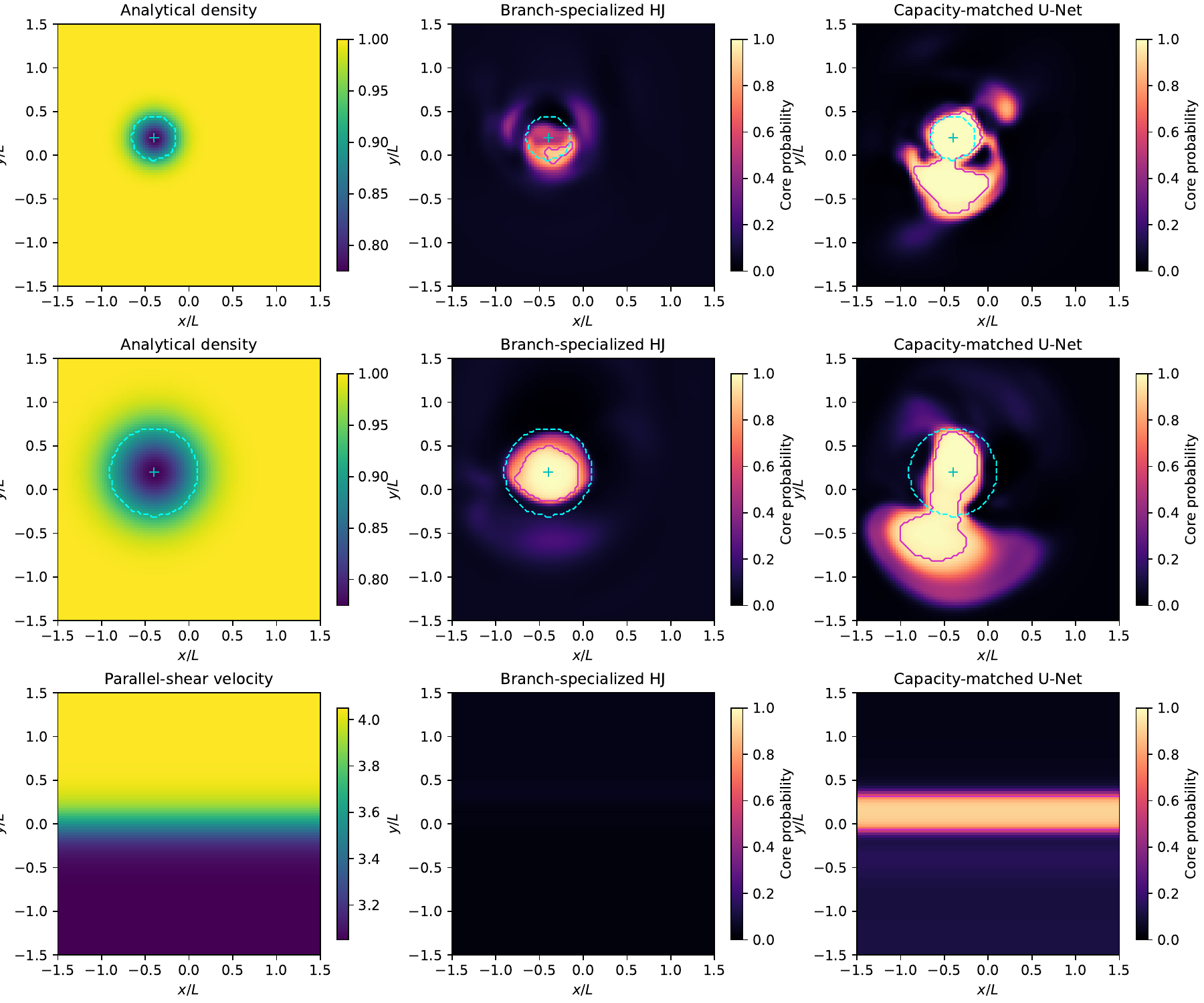}
\caption{Independent vortex and non-vortex controls for the frozen models before shock adaptation. For analytic-vortex panels, cyan marks the exact vortex centre and the prescribed core disk $s\leq1$, whose boundary is the radius of peak tangential speed; magenta is the learned vortex-core boundary. The examples were selected in advance to display three distinct outcomes: a low-Mach miss, a correctly recovered analytic vortex, and a U-Net false positive in parallel shear. The left image in each example shows the corresponding density or velocity field. Because the analytic geometry is prescribed and no threshold is tuned on these cases, this comparison is independent of the CFD weak labels.}
\label{fig:independent-vortex-control}
\end{landscapefigure}
\FloatBarrier

%% file: corrected_target_control_table.tex
\begin{table}[!htbp]
\centering\small
\caption{Equal-additional-budget control after 900 shock-target updates, reported as mean $\pm$ sample SD over three seeds. All four arms use the same corrected shock targets, patch order and validation rules. ``Task-preserving HJ'' updates only the shock branch; ``Shared-decoder U-Net'' allows the shared decoder to change; ``U-Net + retention'' adds a parent-vortex-logit mean-square penalty with fixed weight $\eta=0.2$; and ``Joint HJ control'' updates the full HJ network. Airfoil/cylinder shock and core columns are Dice scores against corrected physics-derived weak references. ``Analytic core'' is Dice against the prescribed isentropic-vortex disk, and ``Centre recovery'' is the fraction of analytic vortices whose known centre is recovered by a predicted object.}
\label{tab:corrected-target-control}
\resizebox{\textwidth}{!}{\begin{tabular}{lrrrrrr}
\toprule
Model & Airfoil shock & Cylinder shock & Airfoil core & Cylinder core & Analytic core & Centre recovery\\
\midrule
Task-preserving HJ & 0.936 $\pm$ 0.004 & 0.939 $\pm$ 0.024 & 0.827 $\pm$ 0.048 & 0.228 $\pm$ 0.020 & 0.672 $\pm$ 0.086 & 0.889 $\pm$ 0.192\\
Shared-decoder U-Net & 0.935 $\pm$ 0.010 & 0.895 $\pm$ 0.022 & 0.083 $\pm$ 0.074 & 0.001 $\pm$ 0.001 & 0.081 $\pm$ 0.066 & 0.472 $\pm$ 0.502\\
U-Net + retention & 0.937 $\pm$ 0.006 & 0.909 $\pm$ 0.013 & 0.775 $\pm$ 0.116 & 0.202 $\pm$ 0.035 & 0.579 $\pm$ 0.047 & 1.000 $\pm$ 0.000\\
Joint HJ control & 0.958 $\pm$ 0.001 & 0.955 $\pm$ 0.021 & 0.382 $\pm$ 0.382 & 0.104 $\pm$ 0.086 & 0.593 $\pm$ 0.192 & 0.861 $\pm$ 0.241\\
\bottomrule
\end{tabular}}
\end{table}

%% file: independent_gasdynamic_validation.tex
\subsection{Independent gas-dynamic reference checks}
\label{sec:independent-gasdynamic}

Analytical shock relations and an empirical bow-shock correlation provide geometric checks that are independent of the training masks. For the diamond airfoil, the oblique-shock relation predicts the angle of each attached leading shock from the incoming Mach number and the surface turning angle. For the circular cylinder, Billig's correlation predicts the detached bow-shock stand-off distance and near-nose shape from Mach number and body radius. These references answer a different question from weak-mask Dice: they test whether the detected front lies where gas dynamics predicts it should. Neither reference curve is fitted, translated or reoriented to improve agreement with the predictions.

\subsubsection{Attached oblique shocks on the continued SU2 fields}
For a calorically perfect gas with ratio of specific heats $\gamma$, the weak attached-shock solution follows from the oblique-shock relation \cite{naca1135}
\begin{equation}
 \tan\theta=2\cot\beta\,
 \frac{M_\infty^2\sin^2\beta-1}
 {M_\infty^2(\gamma+\cos 2\beta)+2},
 \qquad \sin^{-1}(M_\infty^{-1})<\beta<\frac{\pi}{2},
 \label{eq:theta-beta-mach-independent}
\end{equation}
where $\theta$ is the flow turning angle imposed by the surface and $\beta$ is the shock angle measured from the incoming velocity; the lower bound on $\beta$ is the Mach angle. The native wall coordinates confirm a forebody half-angle $\delta=8\dg$. With $M_\infty=3$, $\gamma=1.4$ and the incoming velocity oriented at $+\alpha$ in the stored body coordinates, the upper and lower turning angles $\theta_u$ and $\theta_l$ and the orientations $\phi_u$ and $\phi_l$ of the two shock rays in body coordinates are
\begin{equation}
 \theta_u=\delta-\alpha,\qquad \theta_l=\delta+\alpha,
 \qquad \phi_u=\alpha+\beta(\theta_u),\qquad
 \phi_l=\alpha-\beta(\theta_l),
 \label{eq:oblique-body-orientation}
\end{equation}
and the analytical rays pass through the leading edge as $y=x\tan\phi$. By \cref{eq:oblique-body-orientation}, the $4\dg$ case has two different shocks, with turning angles $4\dg$ and $12\dg$, and at $8\dg$ the upper fore-surface is aligned with the free stream, so only the lower side carries a finite-strength shock.

We first check the CFD fields themselves before comparing any neural boundary with theory. For each leading shock, the native density-gradient ridge is fitted by a straight line using least squares while allowing the line intercept to vary; the fitted orientation gives the ``native'' shock angle in \cref{tab:continued-shock-angles}. The five finite-strength shocks agree with \cref{eq:theta-beta-mach-independent} to within $0.42\dg$, with the largest differences on the lower, more strongly turned sides. This field-level check matters because a detector cannot be expected to agree with an analytical shock ray more closely than the numerical solution itself. The continued fields are used as computed. Their residual histories had not yet reached the convergence criteria adopted for the project (Supplementary Section S1), so the comparison tests the combined CFD-plus-detection pipeline rather than an isolated detector. Two additional consistency checks are reported elsewhere: the body disturbance reaches the far-field boundary, and the shoulder pressure profiles follow the simple-wave trend (\cref{sec:final-multistructure}).
\begin{table}[htbp]
\centering\small
\caption{Field-level validation of the continued SU2 leading shocks against oblique-shock theory. $\beta$ is the shock angle measured from the incoming freestream direction. ``Analytical'' is the weak attached-shock solution of \cref{eq:theta-beta-mach-independent}; ``Native'' is obtained by fitting a straight line to the shock ridge in the solver field without forcing the line through the leading edge; ``Absolute error'' is the magnitude of their angular difference. The $8\dg$ upper side is absent because its surface is aligned with the freestream and therefore has no finite-strength attached shock.}
\label{tab:continued-shock-angles}
\begin{tabular}{llrrr}\toprule
Incidence & Side & Analytical $\beta$ & Native $\beta$ & Absolute error\\\midrule
0 & Upper & 25.6114 & 25.5373 & 0.0741\\
0 & Lower & 25.6114 & 25.6233 & 0.0119\\
4 & Upper & 22.3544 & 22.3560 & 0.0016\\
4 & Lower & 29.2510 & 29.6553 & 0.4043\\
8 & Lower & 33.2886 & 33.7035 & 0.4149\\\bottomrule
\end{tabular}\end{table}

To compare a finite-width raster mask with a one-dimensional analytical ray, we first reduce the mask to one representative trace point per streamwise column. At each abscissa $x_i$, the trace ordinate $y_i$ is the median vertical coordinate of all positive shock pixels inside the prescribed fluid-side window. If a column contains no positive pixel, it remains missing; no point is inserted by interpolation. The signed normal displacement $d_{n,i}$ is the perpendicular distance from that trace point to the analytical ray, and $E_n$ is the mean absolute normal displacement over the $N$ retained columns:
\begin{equation}
 d_{n,i}=\frac{y_i-x_i\tan\phi}{\sqrt{1+\tan^2\phi}},
 \qquad E_n=\frac{1}{N}\sum_{i=1}^{N}|d_{n,i}|.
 \label{eq:analytical-normal-error}
\end{equation}
This extraction never selects the prediction pixel that happens to be closest to theory, which would bias the error downward. A complementary coverage measure samples 301 equally spaced points along the analytical ray and asks what fraction have a positive shock pixel within two raster spacings. Coverage therefore measures how much of the finite reference segment is reached, whereas $E_n$ measures where the recovered band is centered when it is present.

\input{continued_shock_replay}

\subsubsection{Cylinder bow-shock stand-off and shape}
For a circular cylinder of radius $R=D/2$, Billig's cylindrical-body correlation \cite{billig1967} gives a detached-shock reference without any learned mask:
\begin{equation}
 \frac{\Delta_B}{R}=0.386\exp\!\left(\frac{4.67}{M_\infty^2}\right),
 \qquad
 \frac{R_{c,B}}{R}=1.386\exp\!\left[\frac{1.8}{(M_\infty-1)^{0.75}}\right],
 \label{eq:billig-parameters}
\end{equation}
where $\Delta_B$ is the nose stand-off and $R_{c,B}$ the shock curvature radius at its vertex. With the cylinder centred at the origin and flow in the positive $x$ direction, the correlated hyperbola is
\begin{equation}
 x_B(y)=-(R+\Delta_B)+R_{c,B}\cot^2\mu
 \left[\sqrt{1+\left(\frac{y\tan\mu}{R_{c,B}}\right)^2}-1\right],
 \qquad \mu=\sin^{-1}(M_\infty^{-1}),
 \label{eq:billig-shape}
\end{equation}
where $x_B(y)$ is the streamwise position of the shock at transverse coordinate $y$ and $\mu$ is the free-stream Mach angle. At $M_\infty=2.7$, \cref{eq:billig-parameters,eq:billig-shape} give $\Delta_B/D=0.36624$, $R_{c,B}/D=2.32170$ and $\mu=21.738\dg$. Billig's relation is an empirical correlation of experimental shock shapes, so agreement with it checks the Euler solution and the detector together.

All three Euler runs are sampled at $t=8$ on the common 90-cells-per-diameter observation grid used by the frozen transfers, with the fine-grid primitive fields conservatively averaged onto that grid before inference. Within $-1.35\le x/D\le-0.55$ and $|y|/D\le1$, the detected front is the median positive abscissa in each transverse row. The stand-off is $\Delta_{\rm det}=-R-\operatorname{median}(x_i)$ over the eight rows with $|y_i|/D\le0.04$, and the shape discrepancy is the mean of $|x_i-x_B(y_i)|/D$ over 180 rows. All rows contain a trace; none is filled by interpolation, and no shift, curvature fit or correlation-dependent threshold adjustment is applied.

\Cref{tab:independent-billig,fig:independent-billig} report the PE/BIR-path comparison; LocalFront is evaluated separately. The detected stand-off exceeds the correlation by $9.2$--$12.3\%$ for ML-only and by $7.7$--$10.7\%$ for the hybrid, and the mean streamwise shape differences are $0.0188D$--$0.0226D$. The physics-only and hybrid traces coincide in this upstream window, and their stand-off differs from ML-only by at most half a common-grid spacing. Refining from f90 to f180 at CFL 0.20 lowers the ML stand-off excess from 12.25 to 9.22 percent (P/H: 10.73 to 7.70 percent), so the discrepancy is grid dependent and is not a grid-converged detector bias. The shared displacement may reflect CFD and extraction conventions, but the extraction errors are correlated, which prevents us from assigning the discrepancy uniquely to the flow solution or from excluding segmentation error.

\begin{table}[htbp]
\centering\small
\caption{Bow-shock comparison with Billig's empirical cylindrical-body correlation at $M_\infty=2.7$ and $t=8$. Billig predicts the normalized nose stand-off $\Delta_B/D=0.36624$. $\Delta_{\rm det}/D$ is the stand-off measured from the detected trace; ``Difference'' is $100(\Delta_{\rm det}/\Delta_B-1)$, so a positive value means the detected bow shock lies farther upstream from the cylinder than the correlation predicts. ``P/H'' combines the physics-only and hybrid rows because those two traces are identical inside this near-nose measurement window. ``Shape MAE/$D$'' is the mean absolute streamwise distance from the detected trace to the unshifted Billig curve, normalized by diameter.}
\label{tab:independent-billig}
\begin{tabular}{llrrr}
\toprule
Run & Branch & $\Delta_{\rm det}/D$ & Difference (\%) & Shape MAE/$D$\\
\midrule
f90, CFL 0.20 & ML-only & 0.41111 & 12.25 & 0.02258\\
 & P/H & 0.40556 & 10.73 & 0.02194\\
f90, CFL 0.10 & ML-only & 0.40556 & 10.73 & 0.02197\\
 & P/H & 0.40556 & 10.73 & 0.02163\\
f180, CFL 0.20 & ML-only & 0.40000 & 9.22 & 0.02018\\
 & P/H & 0.39444 & 7.70 & 0.01877\\
\bottomrule
\end{tabular}
\end{table}

\begin{landscapefigure}
\includegraphics[width=\linewidth,height=\lsfigureheight,keepaspectratio]{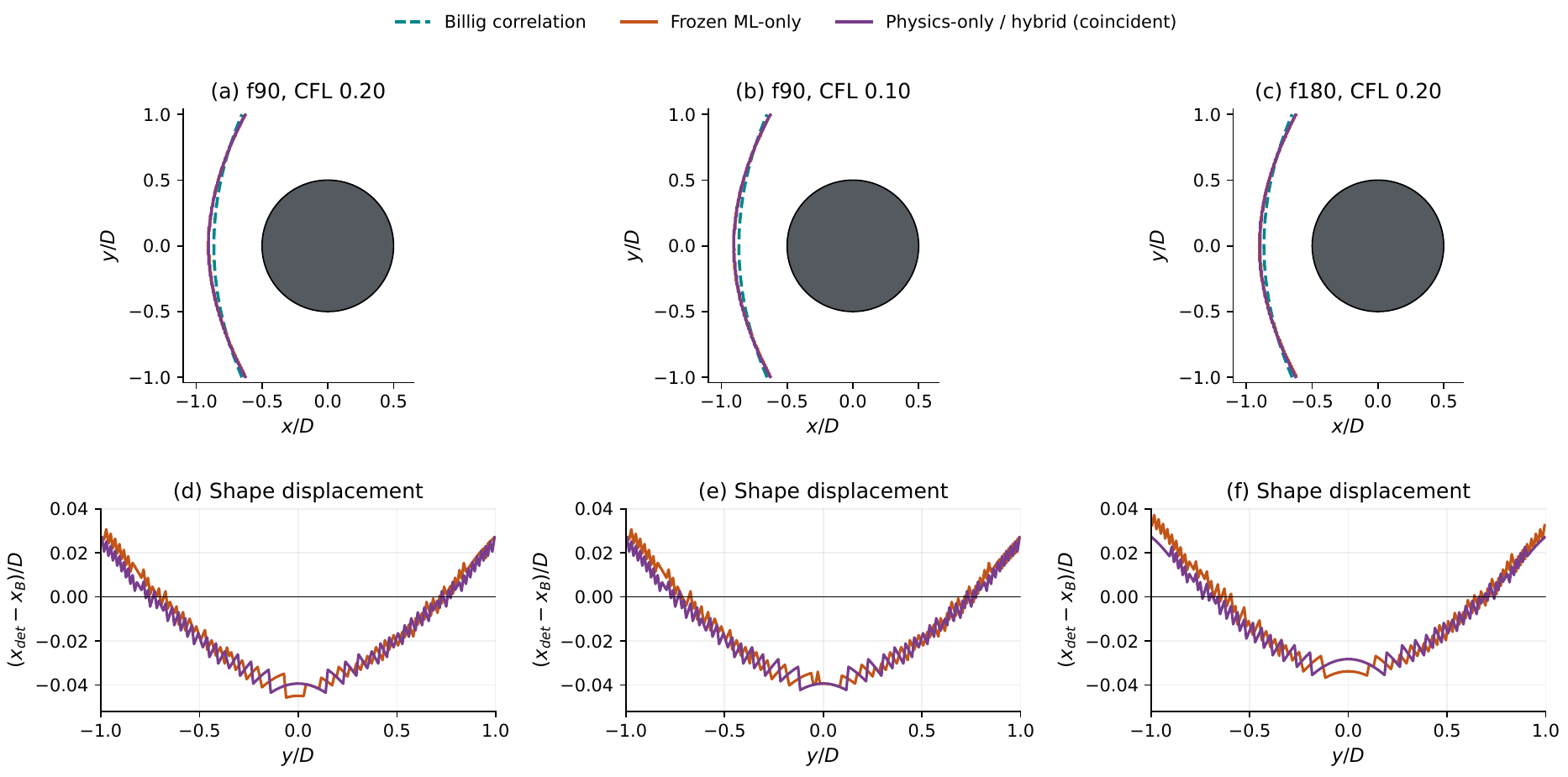}
\caption{Geometric comparison of the detected circular-cylinder bow shock with the unshifted Billig correlation. The top panels show the circular body and only the near-nose portion of the shock used for measurement, with equal $x$ and $y$ scales so curvature is not visually distorted. The Billig curve is the external empirical reference; the learned and physics-supported traces are overlaid without fitting a translation or curvature correction. The lower panels plot signed streamwise displacement $x_{\rm det}-x_B(y)$, so the sign shows whether the detected front is upstream or downstream of the correlation at each transverse location. Physics-only and hybrid traces coincide in this window. Differences among grid/time-step runs are smaller than their common offset from Billig, indicating that both numerical resolution and a systematic CFD/extraction displacement may contribute.}
\label{fig:independent-billig}
\end{landscapefigure}

The two checks constrain shock geometry only. A canonical normal-shock--vortex interaction, such as the one studied by Inoue and Hattori \cite{ref45}, would allow a comparison with a numerically verified post-interaction reference; we do not assume an exact analytical solution for the deformed structures.

%% file: continued_shock_replay.tex
\subsubsection{Frozen learned predictions on the continued SU2 fields}
Four already-trained models are evaluated on the continued SU2 fields without retraining or threshold recalibration. HJ-base is the harmonized joint model before shock adaptation (\cref{sec:v4-study}); HJ-adapted is the task-preserving shock-adapted model (\cref{sec:shock-adaptation-method}); HJ-OP is an orientation-augmented, parent-preserving HJ variant whose shock decoder received 300 additional updates on randomly rotated training patches while a parent-logit penalty discouraged drift from HJ-adapted; and the fourth model is the capacity-matched conventional U-Net of \cref{sec:capacity-unet}. The first initialization seed is used for the base HJ and U-Net, and both HJ adaptations descend from that same base seed. Raw SU2 units have $\rho_\infty=p_\infty=1$ and $U_\infty=3\sqrt{1.4}$, and all four models receive the same freestream-referenced input transform.

The common observation raster has $481\times337$ pixels over $[-0.2,1.8]c\times[-0.7,0.7]c$. Its physical pixel spacing is $h=0.00417c$ in the streamwise direction; this $h$ is the length unit used when a localization tolerance is stated in ``raster spacings.'' Every model sees identical coordinates, primitive fields and solid mask. Full-field inference uses $384\times384$ tiles with a 64-pixel overlap, and connected predicted components smaller than nine pixels are discarded. These settings and the validation-selected probability thresholds are held fixed across the four models.

Localization is evaluated only in the prescribed window $0.10\leq x/c\leq0.40$, outside a $0.015c$ wall band and within $|y|/c\leq0.35$. In each retained streamwise column, the median $y$ coordinate of all positive shock pixels defines one point of the predicted trace. \emph{Coverage} is the fraction of 301 equally spaced analytical-ray samples that have at least one positive prediction within two raster spacings ($0.00833c$). The normal localization error $E_n$ is the mean absolute perpendicular distance from the recovered trace to the analytical ray, as defined in \cref{eq:analytical-normal-error}. Thus coverage answers ``how much of the analytical ray is reached?'' whereas $E_n$ answers ``how far from the ray is the recovered trace?'' The upper fore-surface at $8\dg$ has zero turning and therefore no finite-strength attached shock, leaving five finite shock rays in total.

These fields provide label-independent geometric references, but they were examined during development, so they are not untouched prospective cases. The coverage statistic does not penalize every off-ray false positive, and the displacement is conditional on a recovered trace. A whole-front accuracy claim would need full-region precision, symmetric front distances and sensitivity to raster spacing and tolerance. In particular, the adapted-versus-U-Net difference combines corrected targets and additional updates with the model differences.

\begin{table}[htbp]\centering\small
\caption{Frozen-model localization against the analytical oblique-shock rays of the continued SU2 fields. ``Mean coverage'' is the average fraction of analytical-ray samples lying within two raster spacings of a positive shock pixel over the five finite-strength rays: upper and lower at $0\dg$ and $4\dg$, and lower only at $8\dg$. $E_n/c$ is the mean absolute normal distance from a recovered predicted trace to its analytical ray, normalized by chord, and is computed only where a trace exists. ``Available traces'' reports how many of the five rays produced a measurable trace. One initialization seed is used for each model.}\label{tab:continued-neural-rays}
\begin{tabular}{lrrr}\toprule
Model & Mean coverage (\%) & Mean $E_n/c$ & Available traces\\\midrule
HJ-base & 89.24 & 0.00330 & 5/5\\
HJ-adapted & 100.00 & 0.00296 & 5/5\\
HJ-OP & 100.00 & 0.00298 & 5/5\\
U-Net & 55.81 & 0.00441 & 4/5\\
\bottomrule\end{tabular}\end{table}

As \cref{tab:continued-neural-rays} shows, HJ-adapted and HJ-OP cover all five analytical ray segments at this tolerance, with mean normal displacements of $0.00296c$ and $0.00298c$, less than one raster spacing, and the fitted orientations of their traces lie within $0.6\dg$ of the analytical rays. HJ-base reaches 89\% coverage, and the capacity-matched U-Net 56\% with one ray left without a trace. The protected vortex, geometry and vortex-uncertainty outputs of the two adapted models are bitwise identical to those of HJ-base on all three inputs. Beyond the measurement window, the full-field comparison in \cref{fig:continued-neural-4} shows that the upper front is detected over a shorter extent than the lower one, and that additional responses appear near the wall and inside the shoulder fan, where the expanding flow steepens the density gradient. The corresponding fields at $0\dg$ and $8\dg$ are reported in the Supplementary transfer atlas.
\begin{landscapefigure}
\includegraphics[width=\linewidth,height=\lsfigureheight,keepaspectratio]{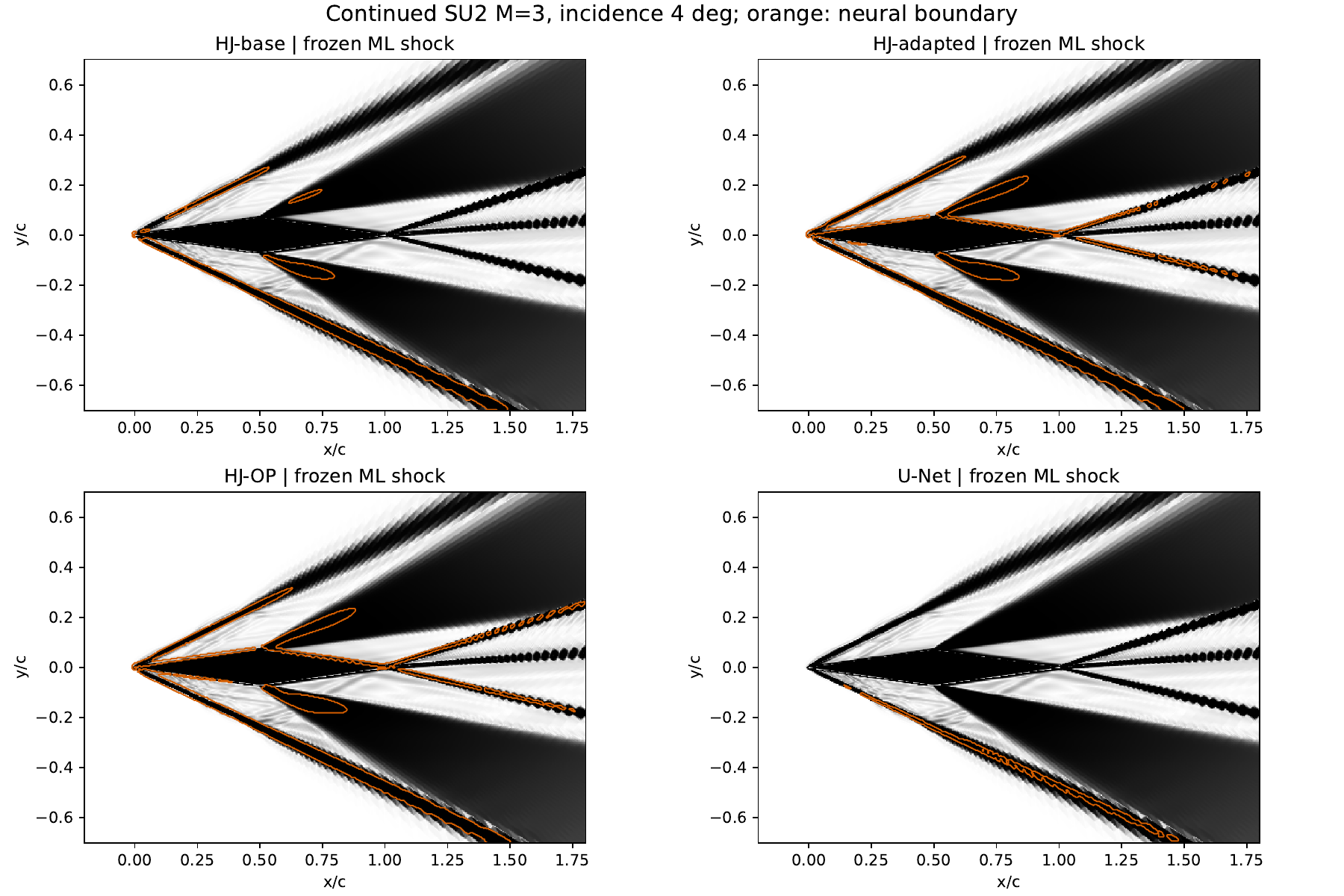}
\caption{Full-field view of frozen shock predictions on the continued SU2 case at $\alpha=4\dg$. Each panel uses the same continued CFD field; only the model changes. Orange curves are boundaries of the ML-only shock mask at each model's fixed validation threshold. HJ-base is the pre-adaptation model, HJ-adapted has the task-preserving shock refinement, HJ-OP adds orientation-augmented shock training with parent preservation, and U-Net is the capacity-matched conventional baseline. The quantitative localization window is only $0.10\leq x/c\leq0.40$ along the leading shocks. Closed orange contours inside the upper and lower shoulder expansion fans are therefore visible failure responses to steepened expansion gradients, not accepted shocks, and they are excluded from the ray-distance measurement. The upper leading shock is also followed over a shorter full-field extent than the lower one.}\label{fig:continued-neural-4}
\end{landscapefigure}

%% file: compact_transfer_atlas.tex
\subsection{Cross-incidence and geometry transfer atlas}
\label{sec:compact-transfer-atlas}
The low-incidence SU2 cases provide the cleanest setting in which to distinguish a broad expanding-flow region from a centred Prandtl--Meyer fan. In \cref{fig:lowfan-a0,fig:lowfan-a4,fig:lowfan-a8}, the left panel is the frozen learned \emph{region} response: it marks pixels whose local state looks expansion-like. The right panel applies a separate geometric decoder. It searches connected learned responses for a convex wall shoulder, verifies local supersonic support, and then represents each accepted fan by two rays: a head ray where the fan begins and a tail ray where the turning is completed. The cyan wedge is therefore an idealized fan object extracted from the learned region, not the outline of every expansion-positive pixel. Analytical Prandtl--Meyer angles are consulted only after this decoding step to measure the angular error. Across the twelve decoded boundaries, the mean and maximum absolute angle errors are $1.25\dg$ and $2.56\dg$. Because the CFD cases were examined during development, this is a diagnostic angle audit rather than an untouched independent fan-segmentation test. At $8\dg$ the learned region also contains thin responses along the zero-turn upper leading side; the shoulder-anchoring rule correctly refuses to turn those isolated responses into another fan object.

\begin{landscapefigure}\includegraphics[width=\linewidth,height=\lsfigureheight,keepaspectratio]{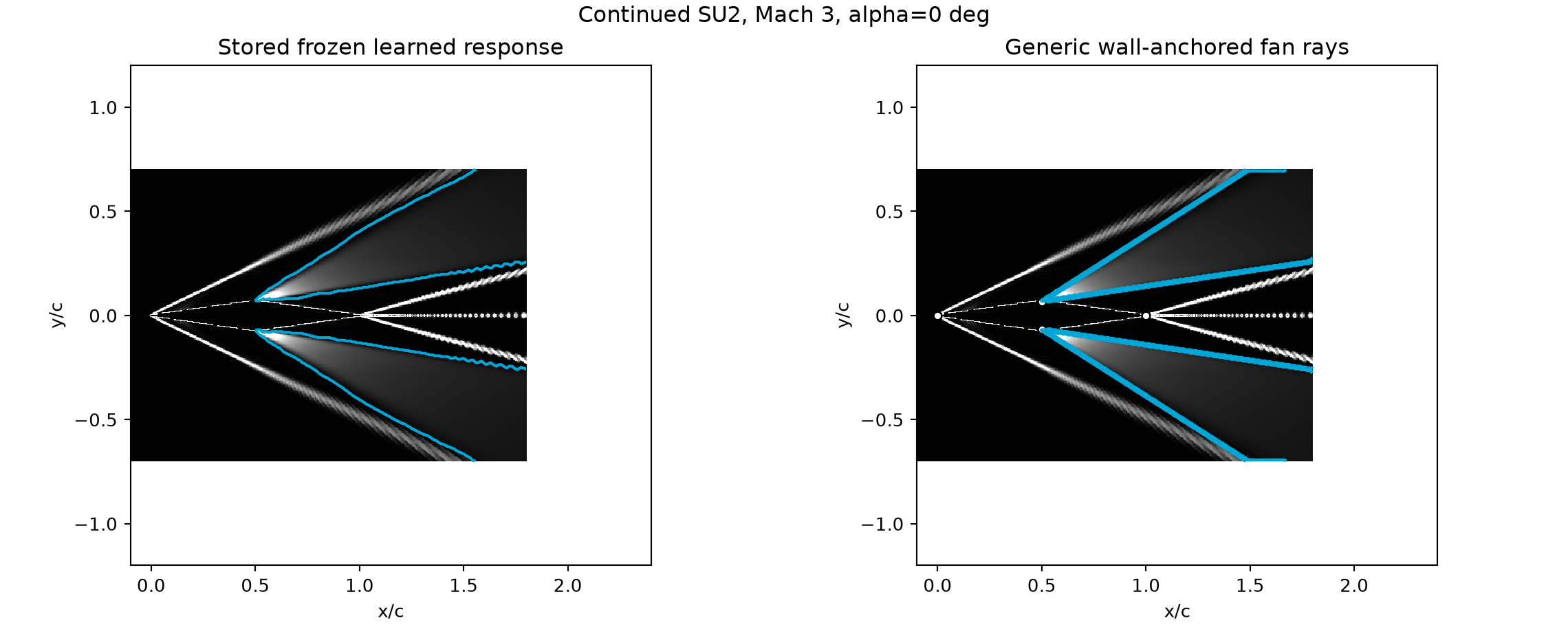}\caption{Continued Mach-3 SU2 field at $\alpha=0\dg$. Both panels use the same numerical-schlieren background and body geometry. Left: cyan outlines the frozen learned expanding-flow region, which is a pixelwise response to acceleration and pressure/density decrease. Right: the same response has been converted into two wall-anchored shoulder-fan objects; each cyan wedge is bounded by a decoded head ray and tail ray originating at the shoulder. These rays summarize a centred fan geometry and should not be read as the outer boundary of the entire gray expanding post-shock layer. White dots identify the wall anchor points used by the geometric decoder.}\label{fig:lowfan-a0}\end{landscapefigure}
\begin{landscapefigure}\includegraphics[width=\linewidth,height=\lsfigureheight,keepaspectratio]{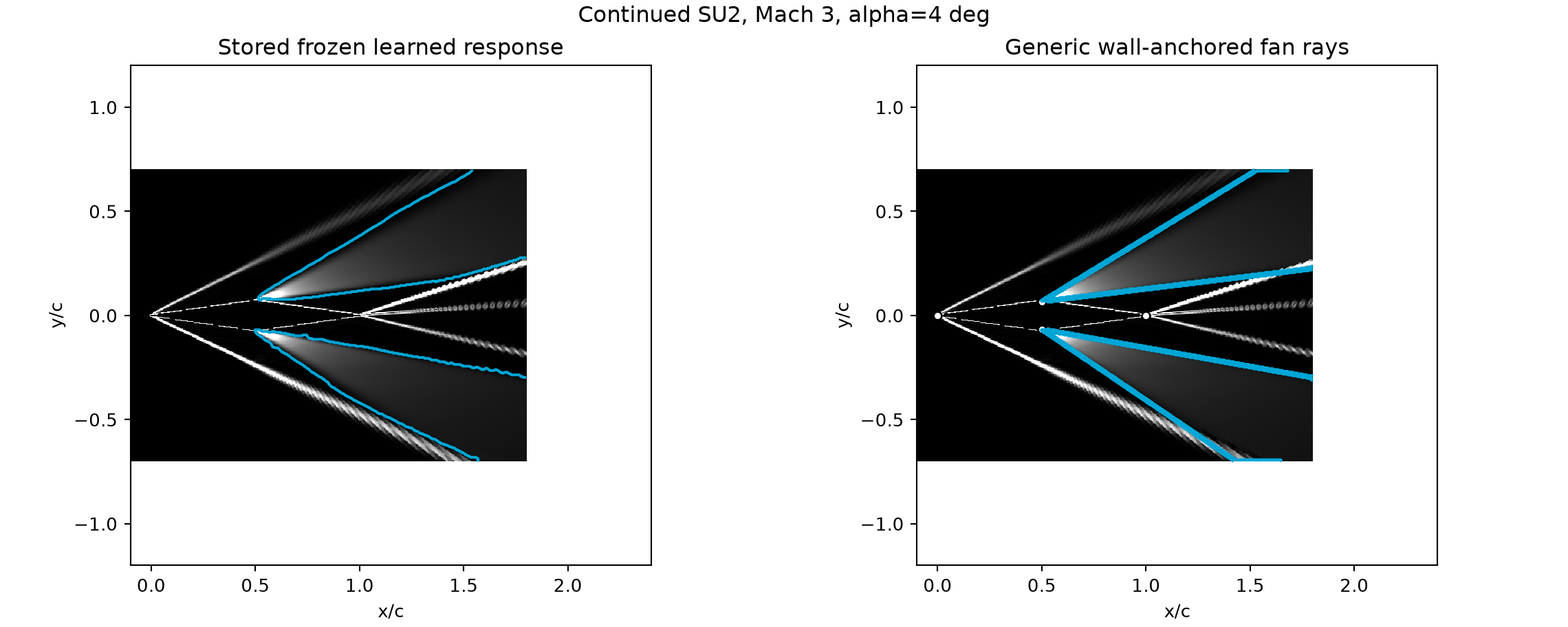}\caption{The same learned-region versus decoded-fan comparison at $\alpha=4\dg$. Cyan in the left panel is the frozen expanding-flow response; cyan wedges in the right panel are the two shoulder-anchored fan objects, each represented by head and tail rays. The decoder is not given the case incidence angle, so the different upper and lower ray orientations arise from the local learned response, wall geometry and supersonic-support test rather than from an imposed $4\dg$ label.}\label{fig:lowfan-a4}\end{landscapefigure}
\begin{landscapefigure}\includegraphics[width=\linewidth,height=\lsfigureheight,keepaspectratio]{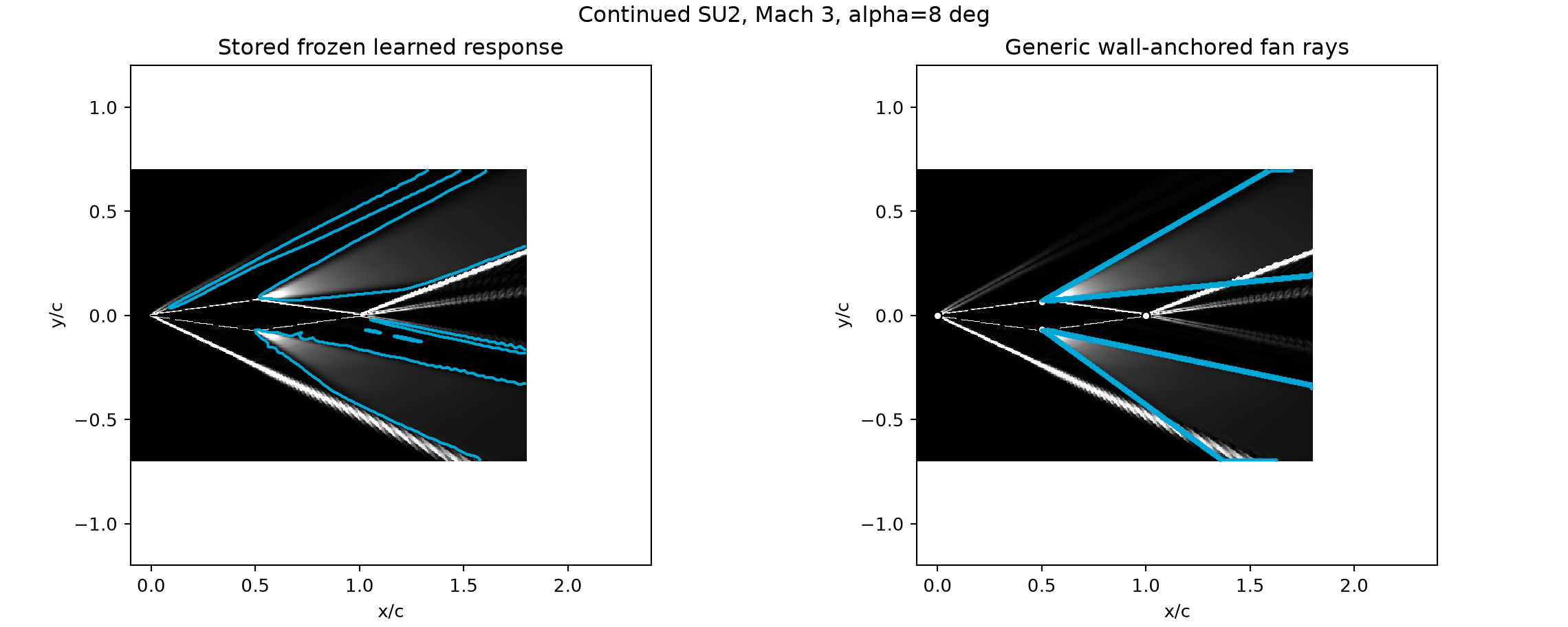}\caption{The same comparison at $\alpha=8\dg$. The upper fore-surface is aligned with the freestream, so the thin cyan responses extending from the upper leading region in the left panel do not correspond to a finite-strength leading-edge fan or shock. They are retained visibly as learned responses rather than hidden. In the right panel the geometric decoder anchors cyan fan wedges only at the physical shoulders, where the wall turns the supersonic stream away from the surface; it does not promote the extra upper-leading responses to fan objects. This panel therefore illustrates why region detection and physically anchored fan interpretation are kept as separate stages.}\label{fig:lowfan-a8}\end{landscapefigure}

The $90\dg$ airfoil and Mach-2.7 ellipse in \cref{fig:transfer-a90,fig:transfer-ellipse} illustrate how the extended output ontology behaves on very different geometries. The models are fixed, but these cases had been examined during development and are therefore demonstrations rather than untouched prospective tests. Orange marks shock/front predictions, purple compact vortex-core candidates, green the connected wake/shear class, and pale blue an expanding-flow \emph{region}. Pale blue is intentionally not labelled as a Prandtl--Meyer fan: a centred fan additionally requires a compatible wall corner and characteristic geometry. The $80\dg$ and $100\dg$ airfoils and the circle and second-ellipse controls are supplied in the Supplementary atlas, including incomplete and ambiguous detections so that transfer failures remain visible.

\begin{landscapefigure}\includegraphics[width=\linewidth,height=\lsfigureheight,keepaspectratio]{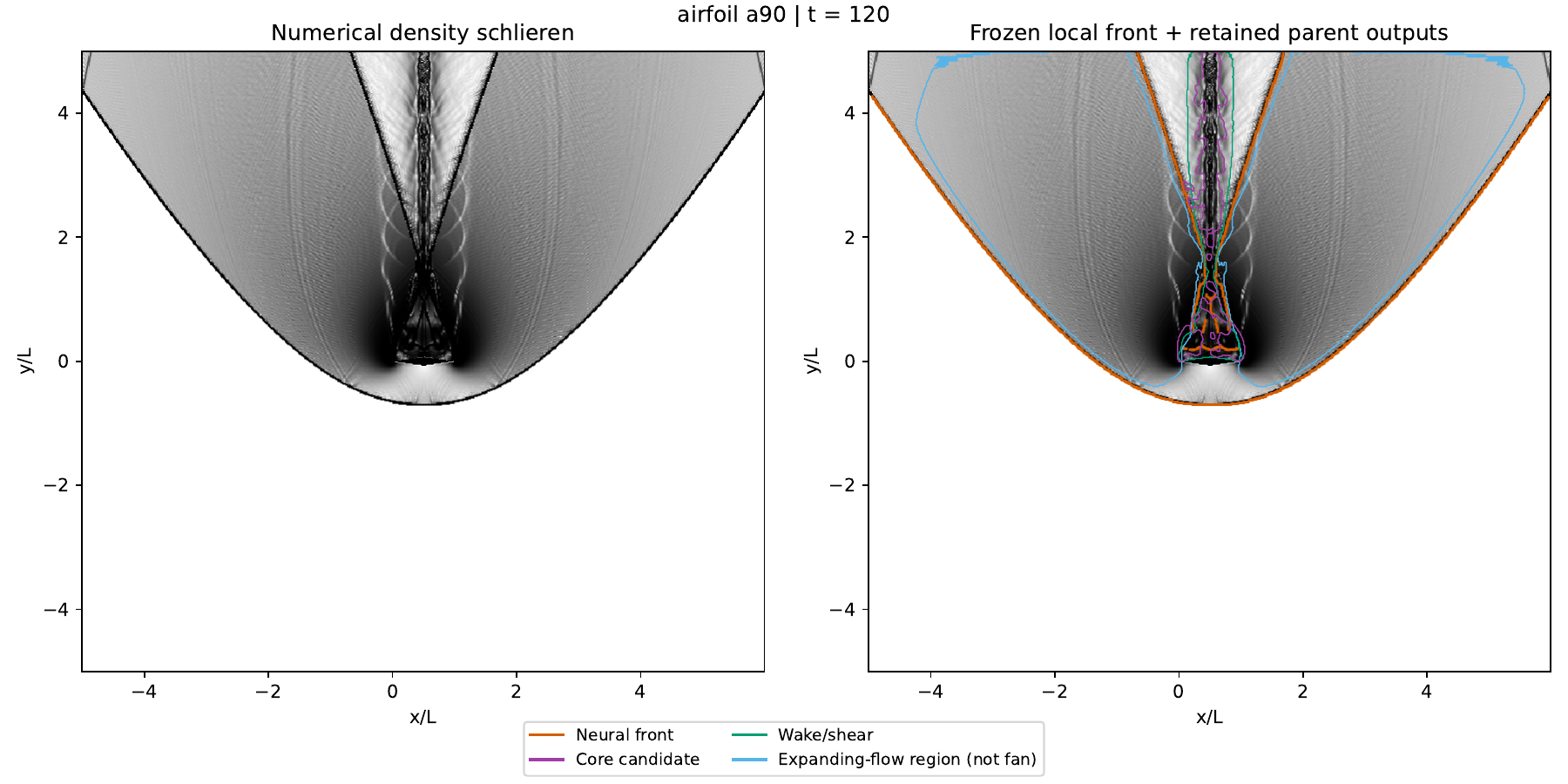}\caption{Fixed-model structure map for the $90\dg$ airfoil at the final stored state. The background is the CFD flow visualization; colored overlays are separate model outputs rather than a mutually exclusive partition. Orange identifies shock/front response, purple compact vortex-core candidates, green wake/shear, and pale blue expanding-flow response. Overlap is allowed where two physical structures coexist. The pale-blue layer is deliberately called an expanding-flow region rather than a Prandtl--Meyer fan because no centred-fan geometry is inferred from the region label alone.}\label{fig:transfer-a90}\end{landscapefigure}
\begin{landscapefigure}\includegraphics[width=\linewidth,height=\lsfigureheight,keepaspectratio]{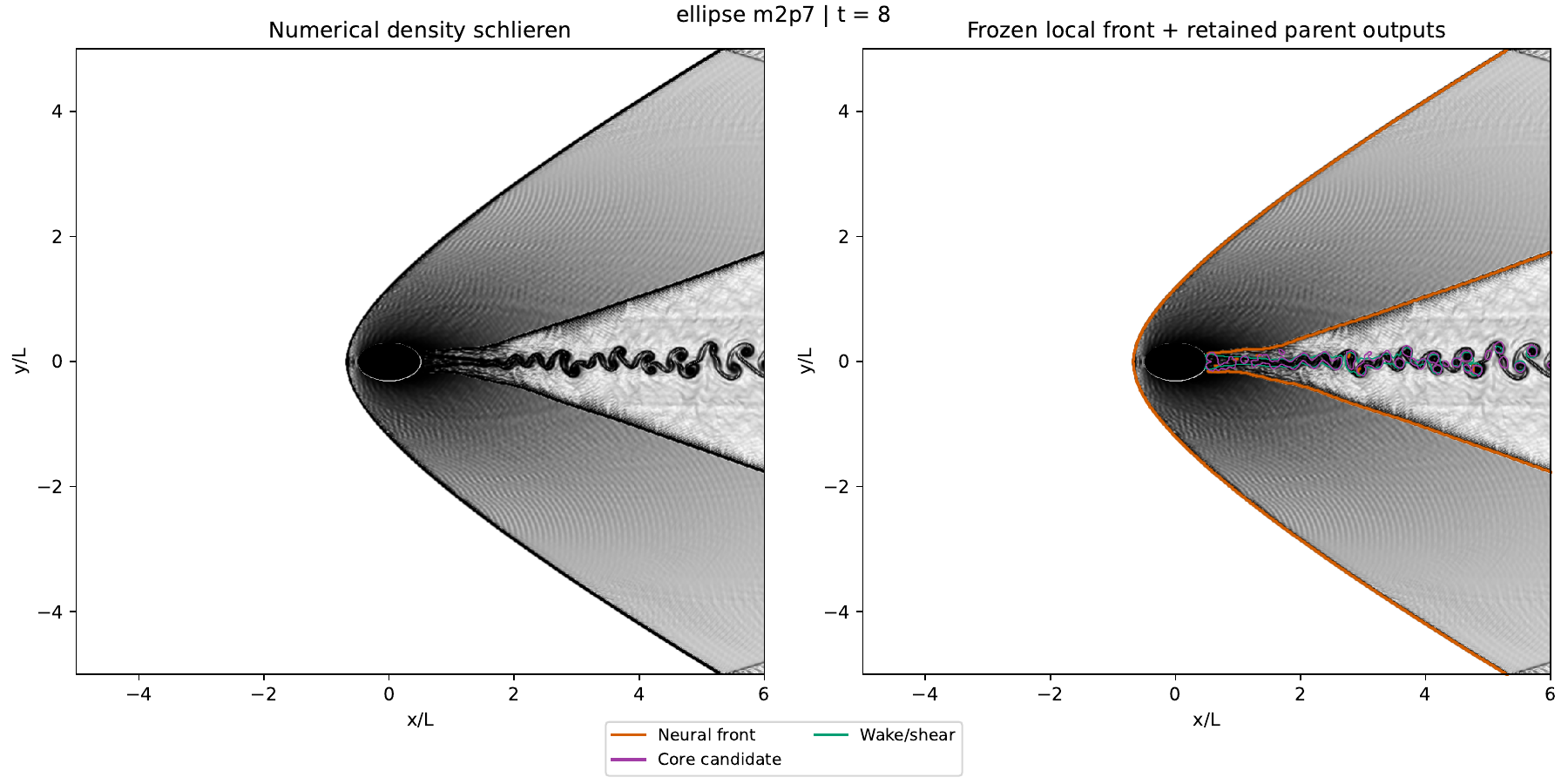}\caption{Fixed-model structure map for the Mach-2.7 elliptical cylinder. Orange traces the detached bow shock and downstream compression/recompression fronts, purple marks compact vortex-core candidates, and green marks the broader connected wake/shear layer. The separate colors emphasize that a compact rotating core and an elongated shear-dominated wake are not treated as the same object. The model is not retrained for this geometry, although the case had been examined during development, so the panel is a transfer demonstration rather than an untouched prospective test.}\label{fig:transfer-ellipse}\end{landscapefigure}
\FloatBarrier

%% file: compact_multistructure.tex
\subsection{Joint physical-structure maps}
\label{sec:final-multistructure}
\begin{figure}[!htbp]
\centering
\includegraphics[width=0.95\textwidth]{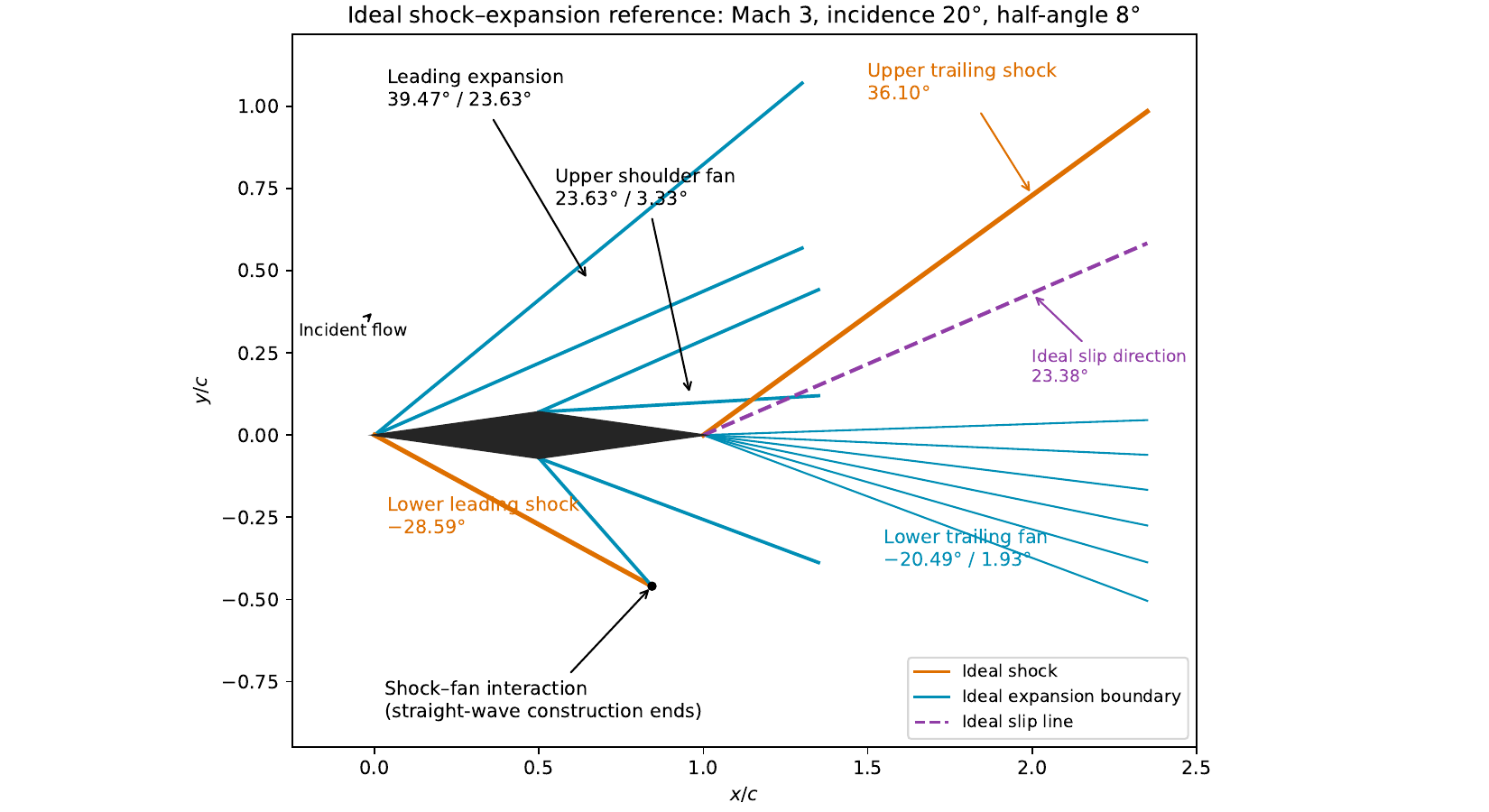}
\caption{Ideal inviscid shock--expansion construction used to interpret the $M_\infty=3$, $\alpha=20\dg$ diamond-airfoil field with half-angle $8\dg$. Solid/dashed wave rays are calculated from the local flow turning: a corner that turns the flow toward itself produces a compression shock, whereas a convex corner that turns the flow away produces a Prandtl--Meyer expansion bounded by head and tail characteristics. The upper leading edge therefore generates an expansion rather than an attached shock. Pressure and direction matching of the downstream streams then require the upper trailing recompression shock, the lower trailing expansion, and the dashed slip direction shown. The lower incident shock is stopped at its first interaction with the fan because the simple piecewise-uniform construction does not predict the curved interacting solution beyond that point. Every colored/dashed element in this schematic is an analytical reference, not a neural prediction or a measured viscous-wake boundary.}
\label{fig:expansion-shock-schematic}
\end{figure}

The shock/core overlays in \cref{fig:v4-repair-airfoil,fig:v4-repair-cylinder} demonstrate task preservation. The $20\dg$ case in \cref{fig:fan-alpha20} addresses a different question: when the network marks an expanding region, does that region correspond to a physically admissible centred Prandtl--Meyer fan? To answer this, the learned region is compared with an ideal inviscid shock--expansion construction for $M_\infty=3$, $\gamma=1.4$ and a diamond half-angle of $8\dg$. A centred fan has two limiting characteristics. The \emph{head} is the first characteristic encountered by the upstream state and the \emph{tail} is the last characteristic after the full turning has occurred. For the upper leading corner their orientations are $39.47\dg$ and $23.63\dg$, while the upper-shoulder fan limits are $23.63\dg$ and $3.33\dg$. Matching the ideal pressures of the streams that meet downstream gives an upper trailing recompression shock at $36.10\dg$, a lower trailing expansion fan from $-20.49\dg$ to $1.93\dg$, and a slip-line direction of $23.38\dg$; \cref{fig:expansion-shock-schematic} shows how those elements are connected. These angles are reference values for an inviscid piecewise-uniform construction. Viscous separation, finite shock thickness and wave interactions can bend or terminate the corresponding features in the CFD field.

Local characteristic checks reject several geometric fan proposals; Supplementary Section S3 gives the per-ray results and the distinction between necessary and sufficient tests.
\begin{landscapefigure}
\includegraphics[width=\linewidth,height=\lsfigureheight,keepaspectratio]{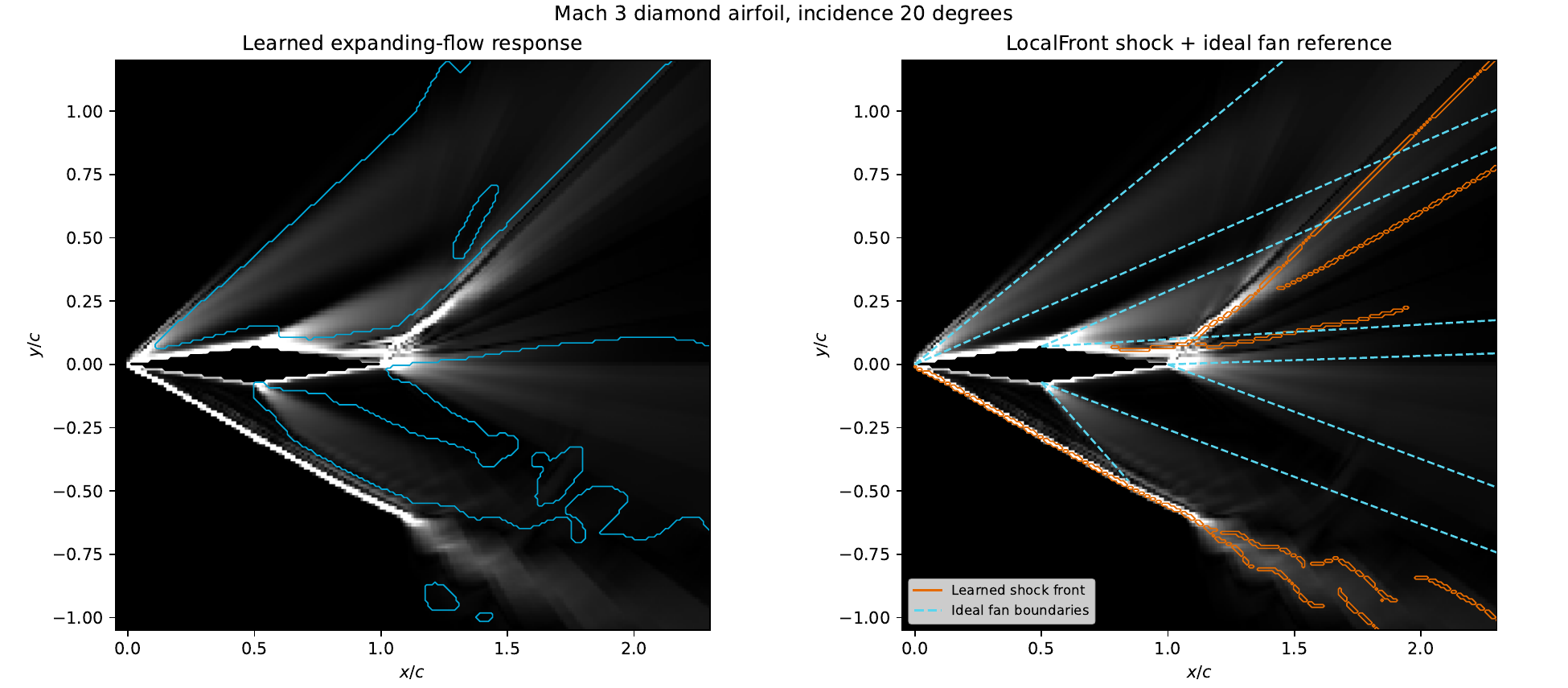}
\caption{Physical interpretation of the continued SU2 field at $\alpha=20\dg$. Both panels use the same CFD density-schlieren background. Left: the frozen network's expanding-flow probability, thresholded at 0.5, shows where the local field resembles expansion; this is a region response and does not by itself define a centred fan. Right: orange is the frozen LocalFront shock/front prediction, while dashed cyan rays are the head and tail characteristics calculated from ideal shock--expansion theory. The cyan rays are analytical references, not neural outputs and not measured CFD boundaries. Straight rays are drawn only to visualize the ideal construction; the lower-shoulder head is terminated where that construction first intersects the incident shock because the post-interaction path is no longer represented by the simple straight-ray solution. The orange segment above the rear upper surface and the fragmented downstream orange trace are unaudited model responses. The learned lower trailing expansion is compatible with the ideal lower fan, whereas isolated responses near the curved shock are not assigned to a fan.}
\label{fig:fan-alpha20}
\end{landscapefigure}

Pressure and flow-direction matching support interpreting the measured upper trailing compression as a recompression branch. An expansion turns and accelerates a supersonic stream while lowering its pressure. Downstream, the stream issued from the opposite side of the body imposes a pressure and direction that cannot, in the ideal construction, be matched by continuing the same isentropic expansion. The compatible ideal solution therefore contains an upper trailing shock. This does not determine how many distinct compression fronts appear in the viscous CFD interaction; that requires local jump evidence. Near an expansion--shock intersection, the front is curved and its local normal rotates from cell to cell, so a one-pixel centerline can look jagged even when the finite-width front is coherent. We therefore interpret the centerline together with the envelope, two-sided thermodynamic samples, grid comparisons and adjacent-time states before attributing oscillations to physical unsteadiness.

The nearly horizontal downstream line was checked separately because a bright schlieren ridge can be produced by several different structures. Pressure and density jumps, a total-pressure proxy, entropy and tangential-velocity changes were sampled across the line at several normal offsets and in adjacent stored states (Supplementary Section S5). Far downstream, the evidence is consistent with a resolved viscous wake/shear layer: it lacks the compressive jump pattern required for a shock, yet its viscous and finite-thickness character also prevents us from certifying it as an ideal inviscid slip line. The immediate interaction region remains mixed. This example is why the manuscript does not identify a structure from schlieren contrast alone.

\FloatBarrier

In the joint maps of this section, each caption specifies the colors and whether a curve is predicted or analytical. Wall, entropy-layer and elongated-core false responses remain visible in the displayed maps; the class definitions do not make such responses physically correct. Extended low-incidence, high-incidence, circular-cylinder and elliptical-cylinder results, including visible failures, are given in the Supplementary material.

%% file: data_availability.tex
\section*{Data and code availability}
The implementation is available under the MIT license at
\url{https://github.com/Ehsan-Roohi/ShockVortexML}.
The source release \texttt{v0.4.0-dev1}, commit
\texttt{\seqsplit{2fb9c1d5b3d70d1a87639273d8b98a1d84e9b251}},
includes the joint-model implementation, one checkpoint and two training examples.
The 195-frame processed learning view and the 2,400 training patches are available
in the versioned research release
\href{https://github.com/Ehsan-Roohi/ShockVortexML/releases/tag/data-v4-20260904}{\texttt{data-v4-20260904}},
with file inventories and SHA-256 checksums. Additional native-precision fields
and model-result supplements are being deposited in the same repository.
Reference masks use the physical
criteria specified in this paper, and versioned manifests identify the datasets,
model parameters and computational records.

The CFD solvers are available at \url{https://github.com/MFlowCode/MFC}
and \url{https://github.com/su2code/SU2}.
The MFC calculations use archived commit
\texttt{\seqsplit{0c9a1d434410175ac483b8d71646455444e3b7eb}}.
Case-generation and cluster-execution scripts are provided at
\url{https://github.com/Ehsan-Roohi/SU2-Diamond-Airfoil-Verification},
including the cylinder branch \texttt{agent/mfc-euler-cylinder-validation}.
The research deposits contain processed fields rather than raw solver restart dumps.

The manuscript source package also supplies the numerical records behind the analytical and empirical reference checks, the continued-field replay, the shoulder-fan audit, the classical-sensor sensitivity sweep, the capacity-matched U-Net comparison and the expansion branch, under \path{validation/}. The three-seed corrected-target aggregate, checkpoint hashes and analytic-vortex summary are supplied under \path{revision_evidence/}; the full prediction arrays remain in the versioned project results because they exceed the Overleaf package budget.

The supplementary movies S1--S10 and their per-movie metadata are distributed under the MIT license in the \href{https://github.com/Ehsan-Roohi/ShockVortexML/releases/tag/movies-localfront-v2-20260909}{ShockVortexML media release movies-localfront-v2-20260909}, separately from the manuscript source.